\documentclass{aa}  

\usepackage{graphicx}
\usepackage{ulem}
\usepackage[mathscr]{euscript}
\usepackage{mathrsfs}
\usepackage{epsfig}
\usepackage{color}
\usepackage{txfonts}
\usepackage{lscape}
\usepackage{natbib}
\usepackage{caption}
\usepackage{soul}
\usepackage{longtable}
\usepackage{captcont}
\usepackage{pdfpages}
\usepackage{amsmath}
\usepackage{color}
\usepackage{soul, xcolor}
\bibpunct{(}{)}{;}{a}{}{,} 

\newcommand{\HI}{H\textsc{i}}

\newcommand{\HIit}{\mbox{H\hspace{0.155 em}{\footnotesize \it I}}}

\newcommand{\MHI}{$M_{\rm HI}$}

\newcommand{\Mz}{$M_{\rm z}$}

\newcommand{\Mstar}{$M_{\star}$}

\newcommand{\Msun}{$M_\odot$}
\newcommand{\kms}{\mbox{km\,s$^{-1}$}}
\newcommand{\nan}{Nan\c{c}ay}
\def\approxlt{\lower.2em\hbox{$\buildrel < \over \sim$}}
\def\approxgt{\lower.2em\hbox{$\buildrel > \over \sim$}}

\newcommand{\FHI}{\mbox{$F_{\rm HI}$}}
\newcommand{\Jykms}{\mbox{Jy~km~s$^{-1}$}}
\newcommand{\kmsMpc}{\mbox{km~s$^{-1}$~Mpc$^{-1}$}}

\newcommand{\Lsun}{\mbox{$L_{\odot}$}}

\newcommand{\Wfifty}{\mbox{$W_{\mathrm 50}$}}

\newcommand{\Lr}{\mbox{$L_{\rm r}$}}

\newcommand{\deltami}{\delta \textrm{M}_{i}}
\newcommand{\Xminus}{\raisebox{2pt}{$\chi$}$_{-}$}
\newcommand{\Xplus}{\raisebox{2pt}{$\chi$}$_{+}$}
\newcommand{\gammau}{\raisebox{2pt}{$\gamma$}}
\newcommand{\epsilonp}{$\epsilon_{+}$}
\newcommand{\epsilonm}{$\epsilon_{-}$}
\defcitealias{butcher2016}{Paper II}
\defcitealias{vandriel2016}{Paper I}
\begin{document}
\setstcolor{red}
    
\offprints{Z. Butcher}
    
\title{Bivariate luminosity-H{\bf \Large I} mass distribution function of galaxies based on the NIBLES survey}
\author{Z. Butcher\inst{1}
        \and
        S. Schneider\inst{1}
        \and
        W. van Driel\inst{2,3}
        \and
        M. D. Lehnert\inst{4}
    }
    
\institute{University of Massachusetts, Department of Astronomy, 619E LGRT-B, Amherst, MA 01003, U.S.A. 
        \email{zbutcher@astro.umass.edu}
        \and
        GEPI, Observatoire de Paris, PSL Universit\'e, CNRS, 5 place Jules Janssen, 92190 Meudon, France  
        \and
        Station de Radioastronomie de \nan, Observatoire de Paris, CNRS/INSU USR 704, Universit\'e d'Orl\'eans OSUC, route de Souesmes, 18330 \nan, France 
        \and    
        Sorbonne Universit\'{e}, CNRS UMR 7095,
        Institut d'Astrophysique de Paris, 98 bis boulevard Arago, 75014 Paris, France  
    }
    
\abstract{We present a new optical luminosity-\HI\ mass bivariate luminosity function (BLF) based on \HI\ line observations from the \nan\ Interstellar Baryons Legacy Extragalactic Survey (NIBLES).  NIBLES sources lie within the local universe (900 $\leq cz \leq$ 12,000 \kms) and were chosen from SDSS DR5 such that the optical luminosity function was sampled as uniformly as possible.  The \HI\ mass function (HIMF) derived from our raw-data BLF, which is based on \HI\ detections only, is consistent with the HIMFs derived from other optically selected surveys in that the low-mass slope is flatter than those derived from blind \HI\ surveys.  However, spanning the entire luminosity range of NIBLES, we identify a highly consistent distribution of the \HI\ gas mass to luminosity ratio (gas-to-light ratio) with a predictable progression in the mean \MHI/\Lr\ ratio as a function of \Lr.  
This consistency allows us to construct plausible gas-to-light ratio distributions for very low-luminosity bins which lie outside the NIBLES sample.  We also identify a $\sim$ 10\% decrease in detection fraction for galaxies fainter than log(\Lr) = 9.25, consistent with the expected decrease due to distance and sensitivity effects.  Accounting for these trends, we reconstruct plausible gas-to-light distributions spanning luminosity bins down to log(\Lr) = 5.25, thus producing a corrected BLF.  This corrected BLF is in good qualitative agreement with optical luminosity-\HI\ mass distributions from the ALFALFA survey and is able to accurately reproduce blind survey HIMFs, lending credibility that this two dimensional optical luminosity-\HI\ mass distribution is an accurate representation of the volume density distribution of galaxies in the local universe.  We also note that our agreement with HIMFs from other surveys is dependent on accounting for all systematic differences such as selection method, Hubble constant and \HI\ flux scale.}
    
    
    \authorrunning{Butcher et al.}    
    \titlerunning{The bivariate luminosity-\HI\ mass distribution}
    \maketitle 
    \section{Introduction}  
    
The optical luminosity function (LF) and the \HI\ mass function (HIMF) are two of the most important and fundamental population tracers of the volume density distribution of galaxies in the universe. They yield clues to both the baryonic and dark matter content as well as their evolutionary histories.  Consequently, both the LF and HIMF are frequently used to constrain galaxy formation models \citep[see, e.g.,][]{benson2003, lu2014}.
    
The LF was the first tracer of volume density to be studied in detail as it was historically easier to detect galaxies at optical wavelengths than in the radio spectrum.  
The first attempts to fit an analytic form to the LF were carried out by, e.g., \citet{Zwicky1957}, \citet{Kiang1961}, and \citet{Abell1965}, but the most successful form was developed by \cite{schechter1976} in which the LF can be characterized by:
\begin{equation}
\Phi(L)dL = \phi^* \Bigg(\frac{L}{L^*}\Bigg)^{\alpha} \text{exp } \Bigg(-\frac{L}{L^*}\Bigg) \ d \ \Bigg(\frac{L}{L^*}\Bigg),
\end{equation}
    
\noindent
where $\phi^*$, $L^*$ and $\alpha$ represent the normalization constant, characteristic luminosity of the ``knee'' of the function and the faint end slope, respectively.  This analytic form has been shown to be a good estimator of the LF across differing environments and many subsequent studies have attempted to analyze how the LF's parameters change as a function of environment and redshift \citep[see, e.g.,][]{felten1985, efstathiou1988, loveday1992,  blanton2001, blanton2003, dorta09, mcnaught2014, loveday2015}.  

The HIMF can be modeled successfully using the same functional form as the LF. It has been analyzed to a somewhat lesser extent due mainly to completeness limitations of past \HI\ surveys.  The HIMF has primarily been studied using several approaches: blind \HI\ surveys \citep[e.g.,][]{zwaan97, zwaan03, kilborn99, kovac2005, martin2010, hoppmann2015} and targeted \HI\ surveys \citep[e.g.,][S05 hereafter]{springob2005a}, and extrapolations from optical surveys using proxies to determine \HI\ content \citep[e.g.,][]{rao1993, solanes1996}.
    
Regardless of the methods used for determining the LF and HIMF, the end goal has always been to determine or analyze the functional form of these two volume tracers independently.  One of the main goals of NIBLES, the \nan\ Interstellar Baryons Legacy Extragalactic Survey \citep[see][Paper I; \citealt{butcher2016}, Paper II]{vandriel2016}, is to determine the correlation between the optical LF and the HIMF -- more specifically, we want to analyze the HIMF and other galaxy properties as a function of optical luminosity.  NIBLES is a 21cm \HI\ line survey at the 100m class \nan\ Radio Telescope (NRT) of 2610 galaxies selected from the Sloan Digital Sky Survey \citep[SDSS; see, e.g.,][]{york00}.  The sample was selected to have radial velocities 900$<$cz$<$12,000 \kms\ and to be distributed evenly over the absolute $z$-band magnitude range of galaxies in the local volume ($\sim$-13.5 to -24), which was used as a proxy for total stellar mass --- see Papers I and II for further details.
    
Here we present the derivation and analysis of the optical luminosity-\HI\ mass bivariate luminosity function (BLF) based on the NIBLES data.  This two dimensional distribution function will provide additional constraints to galaxy formation models and allow more detailed analyses of galaxy populations and evolutionary histories. 
    
In Sect. \ref{sec:NIBLESsample} we summarize the selection criteria of the NIBLES galaxies, in Sect. \ref{sec:BS} we explain our selection of galaxies from the NIBLES sample for the bivariate luminosity and HIMF analysis.  Sect. \ref{sec:method} discusses the method we use to derive our bivariate function and results are discussed in Sect. \ref{sec:results}.  In Sect. \ref{sec:discussion} we compare our resulting HIMFs and $\Omega_{\textrm{\HI}}$ to those of other surveys, both optical and blind \HI, and compare our gas-to-light distributions to those of ALFALFA.  In Sect. \ref{sec:future} we discuss future work and possible ways to improve the accuracy of our BLF and we present our conclusions in Sect. \ref{sec:conclusion}.
    
\section{The NIBLES sample}  
\label{sec:NIBLESsample}
    
The NIBLES galaxy selection criteria are:
\begin{itemize}
\item{Must have both SDSS magnitudes and optical spectrum;}
\item {Must lie within the local volume (900$<$cz$<$12,000 \kms);}
\item {Uniform sampling of each 0.5 magnitude wide bin in absolute $z$-band magnitude, \Mz;}
\item {Preferentially observe the most nearby objects;}
\item {No a priori selection on color.}
\end{itemize}
    
Throughout this paper distances are based on a Hubble constant of $H_{\mathrm 0}$ = 70 \kmsMpc.  Our lower velocity limit of $cz$ = 900 \kms\ was chosen to avoid large distance uncertainties due to peculiar velocities as well as the problems with the SDSS deblender encountered when trying to fit objects with large angular diameters \citep[see, e.g.,][appendix A]{west2005}.
    
We note that the initial targets for NIBLES were selected from SDSS DR5, but the optical data was subsequently updated to DR9 after our observations were concluded to take advantage of the improved SDSS photometric processing pipeline.  NIBLES galaxies were initially selected at random from DR5 until the most densely populated 0.5 mag wide \Mz\ bins contained at least 150 galaxies.  The exceptions were the more sparsely populated highest and lowest magnitude bins, which contained all the SDSS galaxies available at the time.
    
In DR5, association between photometric and spectroscopic targets was made using a position-based match only.  However, this led to (sometimes severe) underestimates of total luminosities for many galaxies due to improper association between the spectroscopic target and a  photometric object that does not contain the most flux from the target galaxy.  This problem has been remedied since DR8, through an additional flux-based association between spectroscopic targets and photometric objects.  (This affected 465 cases marked with flag $F$ in \citetalias{vandriel2016}.)  The effect of updating the photometry of our sample from DR5 to DR9 is that galaxies were systematically shifted to brighter magnitudes.  After this reassessment of total galaxy magnitudes, we still have adequate sampling of the luminosity function over the range ($-23.5 <$ \Mz\ $\la -13.5$) with more than 50 galaxies per 0.5 magnitude wide bin.  In addition to the new flux-based association between photometry and spectroscopy, the post DR8 releases contain re-processed photometry, which affect how we normalize our bivariate luminosity function (BLF).  This will be discussed further in Sect. \ref{sect:uncert}
    
\begin{figure} 
\centering
\includegraphics[width=9cm]{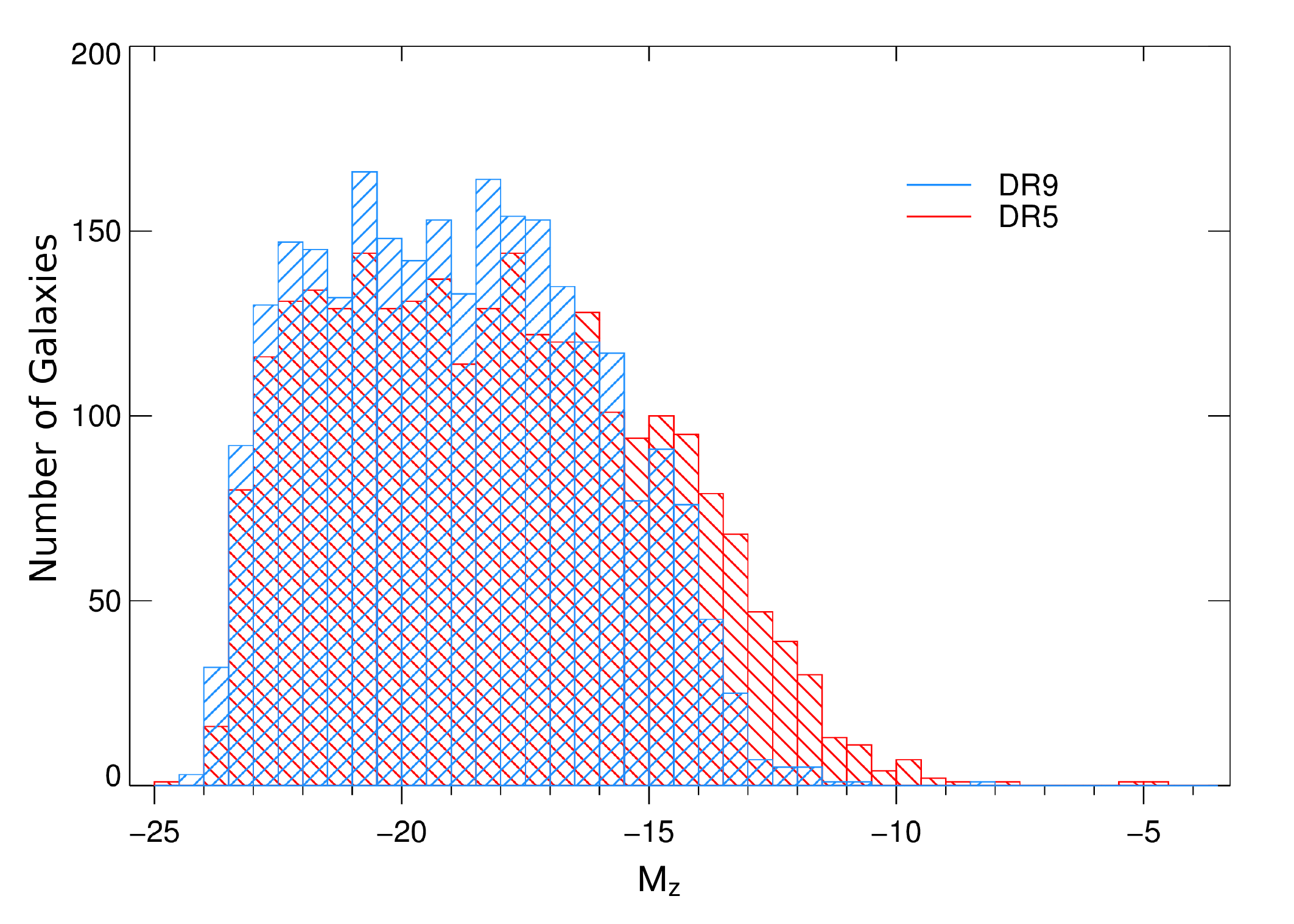}
\caption{\small Difference between the SDSS DR5 and DR9 absolute $z$-band magnitude distributions of the NIBLES sample galaxies.  The DR9 magnitudes were updated using the flux-based association between the spectroscopic target and the photometric object rather than the position-only based match in DR5.  Consequently, there are fewer low-luminosity sources than originally anticipated, as many of their DR5 magnitudes had been (severely) underestimated.}
\label{fig:dr9update}
\end{figure}

NIBLES is aimed to complement other recent and/or ongoing large \HI\ surveys in the local volume, in particular, blind surveys such as ALFALFA \citep[e.g.,][H11 hereafter, and references therein]{haynes2011}.  One advantage of NIBLES, which targets selected objects only, over blind \HI\ surveys is our increased on-source integration time allowing us to reduce the rms noise of our observations.  Each NIBLES source was initially observed for about 40 minutes of telescope time, resulting in a mean rms noise of $\sim$3 mJy.   Clear detections were not observed again, but for weak or non-detections, observations were repeated (as time allowed) resulting in a bimodal rms noise distribution (see Fig. 6 in \citetalias{vandriel2016}) with a mean rms of 2.3 mJy for the latter sources.  
    
Furthermore, we have shown \citepalias{vandriel2016} that the NRT fluxes are on average about 15\% lower than those of the \HI\ flux calibrator galaxies of \cite{oneil04b}.  For the purposes of this analysis, we therefore choose to increase  our \HI\ line flux values by 15\% and re-calculate the \HI\ masses accordingly. We also adjust other \HI\ surveys examined here to the same flux scale (see \citetalias{vandriel2016} for details on flux comparisons).

\subsection{Effects of \HI\ self-absorption} 

Our total \HI\ masses of individual galaxies are calculated using $M_{\textrm{\HI}} = 2.36 \cdot 10^5 \cdot D^2 \cdot F_{\textrm{\HI}}$ where $F_{\textrm{\HI}}$ is the integrated measured line flux (in \Jykms) and $D$ the galaxy distance in Mpc. We make no corrections for possible \HI\ self-absorption (HISA), for two reasons -- lack of consensus on its magnitude, and consistency with other \HI\ surveys.

In principle,  HISA will occur in dense interstellar clouds, and it can decrease the measured integrated \HI\ line flux of an entire galaxy and thus lead to an underestimation of its total \HI\ mass, provided that such clouds are sufficiently abundant.  

Using single-dish radio telescope 21 cm line profiles, e.g. for the HIPASS blind \HI\ survey an HISA correction factor of 1.1 and 1.3 was estimated for the \FHI\ of an average galaxy by Zwaan et al. 1997 and Lang et al. 2002 respectively, whereas Zwaan et al. (2003) concluded that no measurement of the possible effects of HISA could be made as they could not be disentangled from optical dust extinction. 

Using high-resolution interferometric \HI\ images of three Local Group galaxies (M31, M33 and the LMC), \cite{braun2012} derived a HISA correction factor of on average 1.34 to be applied to their measured integrated \FHI\ values.

The HISA correction factor is expected to be the largest for edge-on, gas-rich galaxies (see also Giovanelli \& Haynes 1984), which are commonly found in blind \HI\ surveys. However, the NIBLES sample was selected without a galaxy color bias and hence contains a larger fraction of relatively gas-poor early-type galaxies than a typical blind \HI\ survey.

Furthermore, we want to maintain consistency with recent blind \HI\ surveys such as HIPASS \citep[e.g.,][]{koribalski04, meyer04} and ALFALFA (e.g., \citealt{martin2010}; H11) which did not apply corrections for possible HISA effects.

\section{Bivariate selection criteria}  
\label{sec:BS}
    
As part of our classification process of the NIBLES sample, galaxies were flagged according to various properties (see \citetalias{vandriel2016} for details).  For our bivariate analysis we exclude all galaxies with $C1$, $C2$, $U$ and $R$ flags.  The \HI\ confusion flags $C1$ and $C2$  indicate, respectively, that the \HI\ line profile of a galaxy is either definitely or probably  contaminated by another detection within the telescope beam.  The uncertain SDSS flux flag $U$ indicates galaxies we deemed to either be missing a large portion of the photometric flux due to improper Petrosian radii fitting in the SDSS or due to foreground star contamination.  The resolved \HI\ source flag $R$ indicates sources which may be missing a significant amount of \HI\ flux.   
    
In Papers I and II we defined a marginal detection category which is also excluded here, and those galaxies are treated as non-detections in this paper.
    
After applying the above cuts, our final sample consists of 2256 galaxies. 
    
For this analysis, we use the SDSS $r$-band Petrosian magnitudes rather than the $z$-band magnitudes used to select sources for the NIBLES sample.  We use the $r$-band because the SDSS processing pipeline sets the photometric aperture for each galaxy in this band, which therefore contains the most accurate Petrosian magnitude for a given galaxy.   
    
Additionally, we do not make corrections for Galactic foreground extinction in order to maintain consistency with the luminosity functions from \cite{dorta09} and \cite{blanton2003} which we use in our analysis.  However, since the SDSS galaxies are at high Galactic latitudes, applying an extinction correction has a negligible impact on our results:  the luminosities of the NIBLES galaxies are diminished by only 0.04 dex on average when we apply the extinction corrections from the SDSS catalog data \citep[based on][]{schlegel1998}.  

Note that while our analysis here utilizes non-extinction-corrected Petrosian magnitudes, the magnitudes listed in Papers I \& II are extinction-corrected model magnitudes.

\section{Method}  
\label{sec:method}

The basic method we use is to count the number of galaxies in each logarithmic luminosity and \HI\ mass bin to determine the \HI\ mass distribution of the NIBLES sample.  Then we scale the resulting bivariate distribution to a luminosity function to derive the two-dimensional volume density.  For clarity, we outline the process using the two dimensional stepwise maximum likelihood method (2DSWML) commonly used to determine mass and luminosity functions \citep[see, e.g.,][]{efstathiou1988, zwaan03}.

\subsection{NIBLES \MHI-\Lr bivariate distribution}  
\label{2DSWML}
    
To determine the \MHI\ distribution of the NIBLES sample, we need to determine the probability that a given galaxy will have a particular \HI\ mass.  To do this, we divide the NIBLES sample into $N_{\rm M}$ logarithmic \HI\ mass bins and $N_{\rm L}$ logarithmic luminosity bins of width $\Delta M$ and $\Delta L$ respectively.  The probability that galaxy $i$ will have an \HI\ mass  $M_{\rm HI_i}$ given its luminosity $L_k - \Delta L/2 \le L_i \le L_k + \Delta L/2$ is:

\begin{equation}
p(M_{HI_i} | L_k ) =  \frac{\theta (M_{HI_i} | L_k)}{\int_{0}^{\infty} \theta (M_{HI} | L_k ) dM},
\end{equation}
    
\noindent
where $\theta$ represents an unknown \HI\ mass - luminosity distribution.

The likelihood of a given \HI\ mass for a particular luminosity is simply the product of the probabilities:    \begin{equation}
\mathscr{L} = \prod_{i=1}^{N_g} p_i,
\end{equation}
    
\noindent
where $N_{\rm g}$ is the total number of galaxies in the sample.  To maximize the likelihood, we take the natural logarithm:

\begin{equation}
\begin{aligned}
\ln\mathscr{L} = \sum_{i}^{N_g} \ln p_i = & \sum_{i}^{N_g}\sum_{j}^{N_{M}} B_{ij}(M_i - M_j | L_k) \ln \theta_{jk} \\
& - \sum_{i}^{N_g} \ln
\Big( \sum_{j}^{N_M} H_{ik} \theta_{jk} \Delta M \Big)
\end{aligned}
\label{eqn:lnl}
\end{equation}
    
\noindent set the derivative equal to zero, and solve for $\theta_{jk}$, i.e., the probability that a galaxy will have an \HI\ mass in bin $j$ and a luminosity in bin $k$.  The solution to this equation is straight-forward and need not be discussed here.  The important point to note is that it reduces to:
    
\begin{equation}
\begin{aligned}
\theta_{jk} \Delta M \Delta L = n_{jk}/N_k,
\end{aligned}
\label{eqn:soln}
\end{equation}
    
\noindent where $n_{jk}$ is the total number of galaxies in the $jk$ \HI\ mass and luminosity bin, and $N_k$ is the total number of galaxies in luminosity bin $k$.

Equation \ref{eqn:lnl} is a summation over all galaxies in the sample for each $M_j$ and $L_k$ bin, counting only the galaxies that fall into a particular $M_j$ bin for a given $L_k$.  The $B_{ij}$ denotes whether or not a galaxy is in the $j$ \HI\ mass and $k$ luminosity bin and $H_{ik}$ denotes whether a galaxy is in the $k$ luminosity bin.  
    
When assigning a particular galaxy to a mass bin, we take into account the uncertainty of its \HI\ mass, in the sense that we assign an occupation number that is proportional to the fraction of the galaxy's possible mass range that falls within the bin.  This has the effect of smoothing out the bivariate distribution for the low sample size bins.  Therefore, $B_{ij}$ can take on the following values:
    
\begin{equation}
B_{ij} =
\begin{cases}
1,& \text{if } | \text{ \Xminus} - M_j| \le \Delta M/2 \text{ and } \\
& | \text{ \Xplus} - M_j| \le \Delta M/2 \text{ and } |L_i - L_k| \le \Delta L/2 \\
\text{\gammau}, & \text{ if} \text{ \Xminus} + \Delta M/2 < M_j \text{ and } \\
& \text{ \Xplus} - \Delta M/2 > M_j \text{ and } |L_i - L_k| \le \Delta L/2  \\
(\frac{1}{2} + \text{\epsilonm}) \cdot \text{\gammau}, & \text{if }  |\text{ \Xminus} - M_j| \le \Delta M/2 \text{ and } \\
& |\text{ \Xplus} - M_j| \ge \Delta M/2 \text{ and } |L_i - L_k| \le \Delta L/2 \\
(\frac{1}{2} - \text{\epsilonp}) \cdot \text{\gammau}, & \text{if } |\text{ \Xplus} - M_j| \le \Delta M/2 \text{ and } \\
& |\text{ \Xminus} - M_j| \ge \Delta M/2 \text{ and } |L_i - L_k| \le \Delta L/2 \\
0, & \text{otherwise},
\end{cases}
\label{Bij}
\end{equation}
where \Xminus\ and \Xplus\ are the minimum and maximum \HI\ mass values, \gammau\ $= \Delta M /2 \deltami$ and \epsilonp\ and \epsilonm\ are $(M_j - \text{\Xplus})/\Delta M$ and $(M_j - \text{\Xminus})/\Delta M$ respectively.  
    
$H_{ik}$ takes the following values:
\begin{equation}
H_{ik} =
\begin{cases}
1, & \text{if } |L_i - L_k| \le \Delta L/2 \\
0, & \text{otherwise}.
\end{cases}
\end{equation}
    
By design, this density distribution has
\begin{equation}
\sum_{j=1}^{N_{\rm M}} \theta_{jk} \Delta M \Delta L = \phi (L_k) \Delta L,
\label{eqn:thetaml}
\end{equation}

\noindent where $\phi$ represents the normalized density distribution of the NIBLES sample.  To scale our distribution to a volume density we use the $r$-band luminosity function from \cite[][M09 hereafter]{dorta09} which was also derived from a pre-DR8 SDSS data release, like NIBLES.  We also scaled the M09 luminosity function to match our Hubble constant of 70 \kmsMpc.  
    
We can now calculate the luminosity - \HI\ mass volume density distribution as follows:
    
\begin{equation}
\Phi (M_j, L_k) \Delta M \Delta L= \theta _{jk} \Delta M \cdot \int_{L_k -\Delta L/2}^{L_k + \Delta L/2} \Phi (L) dL.
\label{eqn:HIlumdist}
\end{equation}
    
From this distribution, we can calculate the HIMF and LF by:
\begin{equation}
\sum_{k=1}^{N_L} \Phi (M_j, L_k) \Delta M \Delta L = \Phi (M) \Delta M  
\label{eqn:HIMFLF}
\end{equation}
and
\begin{equation}
\sum_{j=1}^{N_M} \Phi (M_j, L_k) \Delta M \Delta L = \Phi (L) \Delta L.
\label{eqn:LF}
\end{equation}
    
Since our \HI\ masses are distributed according to luminosity (due to our selection criteria), we must treat the luminosity uncertainty differently than the \HI\ mass uncertainty.  Due to our normalization to a luminosity function, any luminosity uncertainties applied in the method described above would be normalized out in Eqn. \ref{eqn:HIlumdist}.  We therefore apply the luminosity uncertainties via a bootstrapping method described in Sect. \ref{sect:uncert} and assign a final value to each $\theta_{j,k}$ bin from the mean values of the bins in that bootstrapped distribution.

\subsection{Differences between SDSS DR5 and DR9 luminosities}  
\label{sect:uncert}
    
We calculate uncertainties in the BLF using two separate methods: the standard deviation of the binomial distribution, to address sampling errors, and a bootstrapping method to address galaxy luminosity uncertainties. The standard deviation of the binomial distribution for each bin in our BLF is given by:

\begin{equation}
\sigma = \sqrt{\frac{\theta_{jk}(1-\theta_{jk})}{n}}
\label{sdbinomial}
\end{equation}
\noindent where $n$ is the total number of galaxies in each luminosity bin.
    
\begin{figure}  
\centering
\includegraphics[width=9cm]{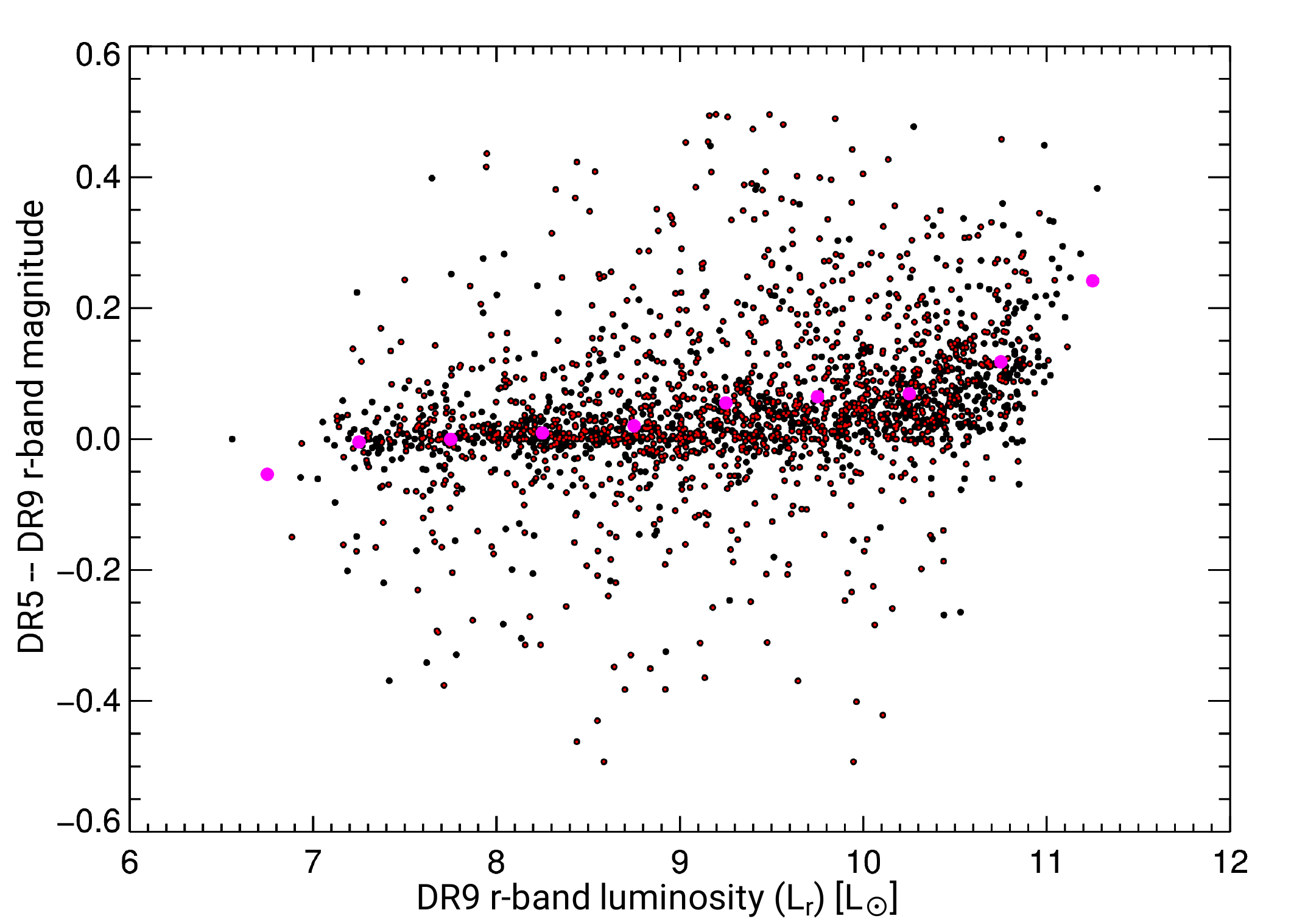}
\caption{\small Difference between SDSS DR5 and DR9 $r$-band apparent Petrosian magnitude as a function of DR9 $r$-band luminosity for all NIBLES galaxies.  Red dots indicate \HI\ detections, black dots non-detections and magenta dots indicate the mean value of the magnitude difference in each 0.5 dex wide bin in \Lr.}
\label{fig:DR5DR9Lr}
\end{figure}

The bootstrap method addresses errors from the uncertainties in galaxy luminosities in the SDSS database.  The listed Petrosian magnitude uncertainties in the SDSS data releases are on average a factor of four lower than the uncertainties in the \HI\ masses.  However, these values will be underestimated as the listed uncertainty is only a function of the flux uncertainty within the aperture fitted to each galaxy by the processing pipeline \citep[see][]{york00}, but does not account for uncertainties introduced by fitting errors.  As the SDSS does not provide estimates for the accuracy of this fit, we estimate it ourselves by comparing the scatter of the difference between  DR5 and DR9 Petrosian magnitudes.  For each DR9 photometric source from the NIBLES catalog, we pulled the nearest corresponding primary photometric source from DR5 for comparison.  Every source in the NIBLES sample was visually checked to ensure that the SDSS DR9 photometric object was centered on the galaxy and contained an adequate Petrosian radius to account for all, or most, of the flux (see \citetalias{vandriel2016} for details). The difference between DR5 and DR9 apparent $r$-band Petrosian magnitudes as a function of DR9 $r$-band luminosity is shown in Fig. \ref{fig:DR5DR9Lr}.
    
There is an increasing discrepancy between the DR5 and DR9 magnitudes with increasing $r$-band luminosity, as indicated by the magenta dots which represent the mean value of DR5 $-$ DR9 magnitudes in each 0.5 dex wide bin in \Lr.  The standard deviation within each bin is fairly consistent at about 0.1 mag, but slightly lower in the two lowest luminosity bins.  The scatter in this relationship is about three to four times larger than the uncertainties listed in the SDSS database. The uncertainties in luminosity as indicated by this scatter would place them in the same range as the \HI\ mass fractional uncertainties.
    
The relationship shown in Fig. \ref{fig:DR5DR9Lr} provides us with both a measure of typical total photometric uncertainties and the offsets between the SDSS DR5 and DR9 data.  The latest derivation of an SDSS luminosity function is from M09 which uses photometry from DR6 (which uses the same photometric processing pipeline as DR5).  Since the more luminous galaxies have a greater difference between DR9 and DR5 magnitudes, we must take this offset into account when normalizing NIBLES DR9 data to the M09 luminosity function.  To accomplish this, we create a bootstrapped sample as follows:  we convert the $y$-axis from Fig. \ref{fig:DR5DR9Lr} to differences in luminosity and bin the data in 0.02 dex wide bins.  Then, for each galaxy in our bivariate sample, we randomly assign a luminosity offset which is drawn from this luminosity difference distribution within its DR9 luminosity bin.  This gives us a new set of luminosities from which we derive a new BLF.  We repeat this process 100 times, generating a new BLF each time.  We then calculate the standard deviation in each $\theta_{j,k}$ bin across our bootstrapped sample.  The final, total uncertainty for each $\theta_{j,k}$ bin is the quadrature sum of the standard deviations of the bootstrapped sample and the binomial distribution in each bin.  Additionally, each $\theta_{j,k}$ bin itself is assigned the median value of our bootstrapped sample, as mentioned in the previous section.
    
\begin{figure}  
\centering
\includegraphics[width=9cm]{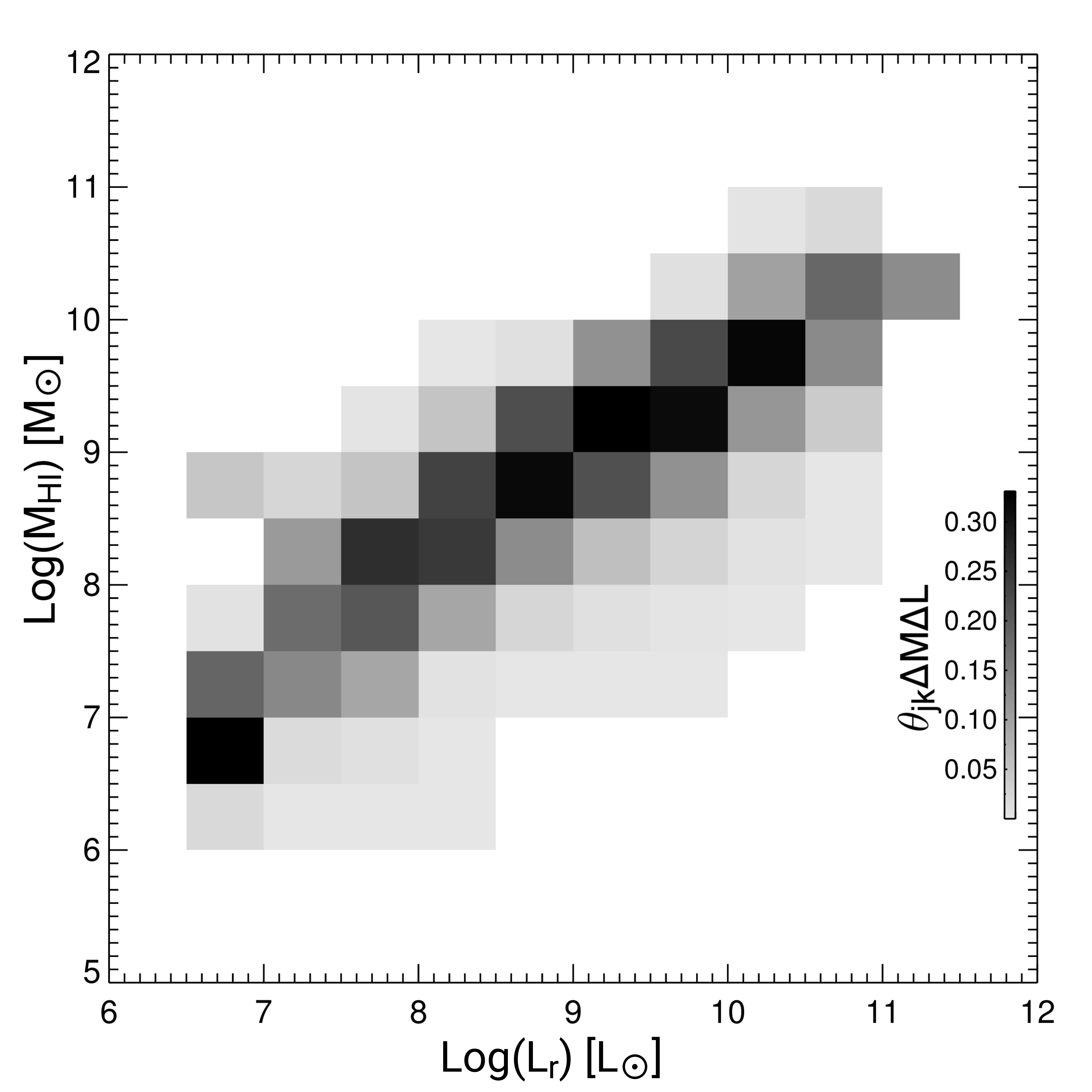}
\caption{\small Two dimensional \HI\ mass - $r$-band luminosity distribution of the NIBLES sample, derived using only \HI\ detections.  The distribution is shaded according to the fraction of galaxies that have a particular log(\MHI) for a given \Lr, see the legend.}
\label{fig:HILrdensnib}
\end{figure}
    
\section{Results}  
\label{sec:results}
    
Fig. \ref{fig:HILrdensnib} shows the 2D \HI\ mass - $r$-band luminosity distribution of the NIBLES sample ($\theta_{jk}$ from Equation \ref{eqn:soln}) binned in 0.5 dex intervals in both log(\Lr) (in \Lsun) and log(\MHI) (in \Msun).  The distribution is shown such that the sum over each luminosity bin is equal to one.  The plot shows (see also Fig. \ref{fig:ML_ratio}) that the galaxies in the highest-density bins have log(\MHI/\Lr) ratios that lie within about 0.5 dex of zero, with this ratio becoming more negative with increasing luminosity.  The highest and lowest \HI\ mass bins, being on the edges of the distribution, are not fully sampled.  In particular the log(\Lr) = 6.75 and 10.75 bins only contains four and six galaxies respectively, with their means in log(\MHI) being 6.8 and 10.6.  The density in the log(\MHI) = 6.25 bin is much lower than in the next highest bin,  which is due entirely to fractional occupation numbers being assigned in this bin as a result of our method described in Sect. \ref{sec:method}.
    
\begin{figure}  
\centering
\includegraphics[width=9cm]{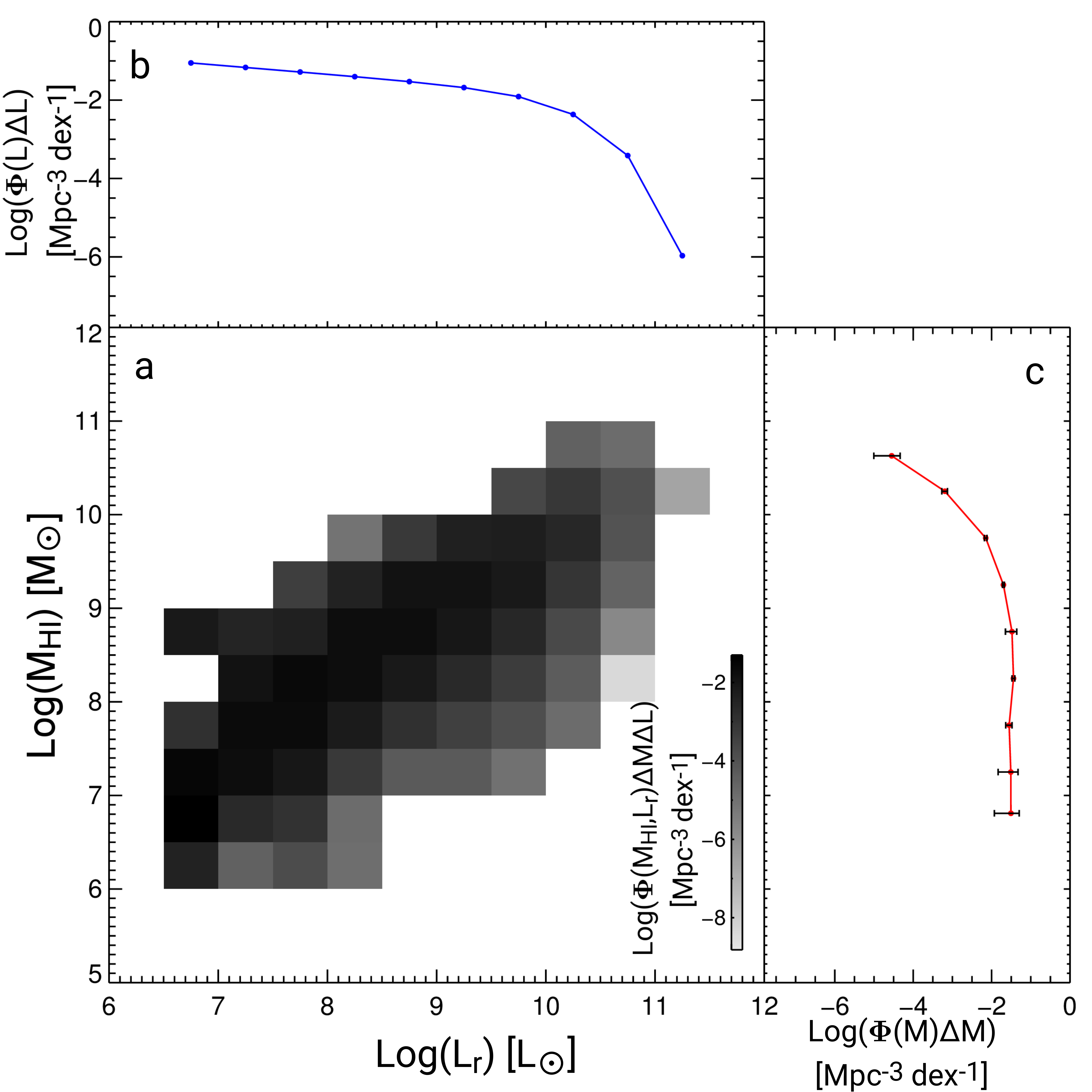}
\caption{\small Two-dimensional bivariate \HI\ mass--\Lr\ distribution of the NIBLES sample, derived using only \HI\ detections.  Main panel $a$: log($\Phi$(M)$\Delta$M) as a function of both \HI\ mass (in \Msun) and \Lr (in \Lsun).  The upper panel $b$ is the summation of the main panel over \HI\ mass, which reproduces the input M09 luminosity function, while right panel $c$: HIMF, the summation of the bivariate distribution over luminosity.  Since the highest \HI\ mass bin in this distribution (log(\MHI) = 10.75) is not fully sampled, we plotted its position in the HIMF (panel $c$) at the mean value of the measured \HI\ masses contained within the bin rather than at its normal \MHI\ midpoint.  We did not plot the point for the lowest \HI\ mass bin (log(\MHI) = 6.25) in the HIMF ($c$) since it was populated with partial occupation numbers only, due to the relatively high fractional \HI\ mass uncertainty of these sources.}
\label{fig:HILrspacedens}
\end{figure}

Normalizing the bivariate distribution to the M09 luminosity function yields the two-dimensional luminosity-\HI\ mass distribution function shown in Fig. \ref{fig:HILrspacedens}.  Panel $a$ shows our resulting bivariate \MHI-\Lr\ distribution.  The input LF from M09 is shown in panel $b$ and the resulting HIMF obtained by summing over luminosity is shown in panel $c$.  Similarly to Fig. \ref{fig:HILrdensnib}, the lowest \HI\ mass bin (log(\MHI) = 6.25) is not completely sampled.  We therefore do not plot this bin in the corresponding HIMF of panel $c$ of Fig. \ref{fig:HILrspacedens}.
    
Values for the mass distribution function from panel (a) in Fig. \ref{fig:HILrspacedens}  are listed  in Table \ref{tab:BLF} as log($\Phi$(\MHI,\Lr)$\Delta M$ $\Delta L$) in units of Mpc$^{-3}$ dex$^{-1}$, together with their fractional uncertainties.
\onecolumn
\begin{table}
\caption{$r$-band luminosity-\HI\ mass distribution function from panel $a$ in Fig. \ref{fig:HILrspacedens}}
\small
\renewcommand{\tabcolsep}{1pt}    
\begin{tabular}{ |l||c|c|c|c|c|c|c|c|c|c|}
\hline
& \multicolumn{10}{|c|}{log(\Lr) bin [\Lsun]} \\
\hline
log(\MHI) bin [\Msun] & 6.75 & 7.25 & 7.75 & 8.25 &8.75 & 9.25 & 9.75 & 10.25 & 10.75 & 11.25 \\
\hline
\hline
6.75&-1.53 $\pm$ 0.29&-2.94 $\pm$ 0.32&-3.22 $\pm$ 0.27&-5.08 $\pm$ 2.11&&&&&& \\
\hline
7.25&-1.78 $\pm$ 0.42&-2.04 $\pm$ 0.11&-2.32 $\pm$ 0.09&-3.45 $\pm$ 0.27&-4.50 $\pm$ 0.78&-4.49 $\pm$ 0.64&-5.21 $\pm$ 1.16&&& \\
\hline
7.75&-3.12 $\pm$ 2.39&-1.92 $\pm$ 0.09&-1.98 $\pm$ 0.06&-2.44 $\pm$ 0.08&-3.13 $\pm$ 0.15&-3.65 $\pm$ 0.23&-4.10 $\pm$ 0.31&-5.11 $\pm$ 0.60&& \\
\hline
8.25&&-2.13 $\pm$ 0.13&-1.87 $\pm$ 0.05&-2.01 $\pm$ 0.05&-2.42 $\pm$ 0.06&-2.92 $\pm$ 0.10&-3.48 $\pm$ 0.15&-4.52 $\pm$ 0.30&-8.63 $\pm$ 12.15& \\
\hline
8.75&-2.38 $\pm$ 1.09&-2.79 $\pm$ 0.28&-2.58 $\pm$ 0.14&-2.03 $\pm$ 0.05&-2.03 $\pm$ 0.03&-2.35 $\pm$ 0.05&-2.82 $\pm$ 0.06&-4.00 $\pm$ 0.16&-5.99 $\pm$ 0.67& \\
\hline
9.25&&&-3.62 $\pm$ 0.44&-2.69 $\pm$ 0.11&-2.19 $\pm$ 0.05&-2.16 $\pm$ 0.03&-2.41 $\pm$ 0.04&-3.31 $\pm$ 0.07&-4.81 $\pm$ 0.15& \\
\hline
9.75&&&&-5.31 $\pm$ 2.28&-3.46 $\pm$ 0.23&-2.60 $\pm$ 0.06&-2.56 $\pm$ 0.05&-2.86 $\pm$ 0.04&-4.29 $\pm$ 0.08& \\
\hline
10.25&&&&&&&-3.86 $\pm$ 0.23&-3.36 $\pm$ 0.07&-4.15 $\pm$ 0.06&-6.85 $\pm$ 0.46 \\
\hline
10.75&&&&&&&&-4.70 $\pm$ 0.39&-5.07 $\pm$ 0.20& \\
\hline
\end{tabular}
\label{tab:BLF}
Volume density values for the data from panel $a$ in Fig. \ref{fig:HILrspacedens}, i.e., log($\Phi$(\MHI,\Lr)$\Delta M$ $\Delta L$) in Mpc$^{-3}$ dex$^{-1}$.  The listed uncertainties are fractional.
\end{table}
\twocolumn

\begin{figure}  
\centering
\includegraphics[width=9cm]{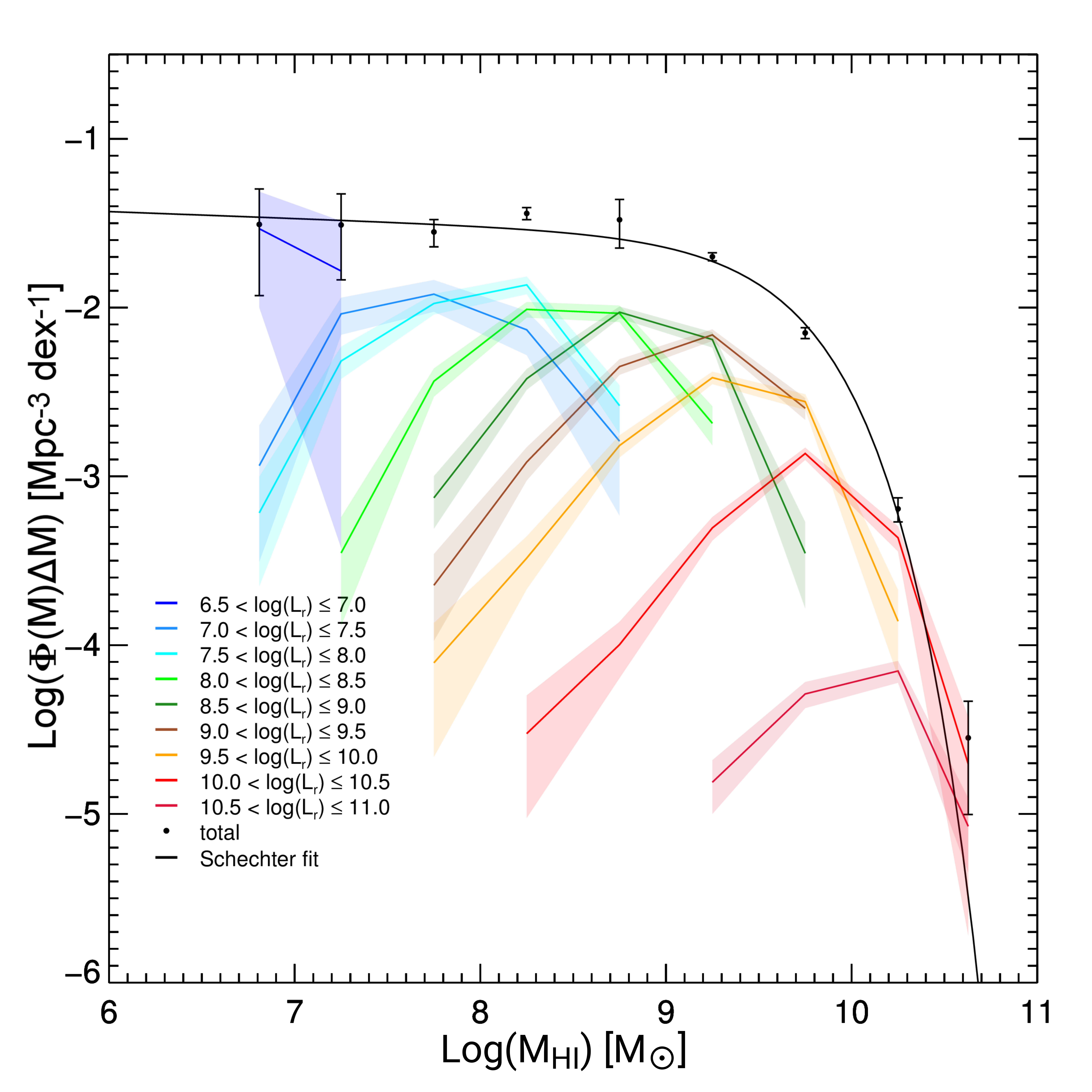}
\caption{\small HIMF separated by luminosity: log($\Phi$(M)$\Delta$M) in units of Mpc$^{-3}$ dex$^{-1}$ as a function of \HI\ mass in each luminosity bin.  For clarity, we only show bins containing more than one galaxy to eliminate the partially sampled bins with large uncertainties.  The black points are the sum of each \HI\ mass for the corresponding luminosity bins.  Uncertainties for each luminosity bin are shown as shaded regions around each mass function, with the total quadrature sum shown as error bars on the HIMF.  The black line is the Schechter fit to the HIMF.}
\label{fig:lumbins_r_np}
\end{figure}
    
In Fig. \ref{fig:lumbins_r_np} we show the HIMF from panel $c$ of Fig. \ref{fig:HILrspacedens} broken down into its corresponding fully-sampled luminosity bins.  The low mass end shows clear flattening, in contrast to blind \HI\ surveys such as \citet[][Z03 hereafter]{zwaan03} and H11.  This flattening is the result of two effects: (1) \HI\ detectability limits for low-luminosity galaxies at $V >$ 900 \kms, and (2) low-luminosity but high \HI-mass galaxies that were not sampled due to being at or below the detection limit of the SDSS.  We explore these two effects in the following two subsections.
    
\subsection{\HIit\  sensitivity limits}  
\label{sec:HI_sens_lim}
The first  cause of the HIMF faint-end flattening is that the NIBLES sample is optically selected with a recession velocity of $V_{\textrm{hel}}$ > 900 \kms, in contrast to the abovementioned \HI-selected surveys, Z03 and H11, which do not have a lower velocity limit -- the former is about five times less sensitive and the latter has about the same sensitivity as NIBLES (see \citetalias{vandriel2016}).  Our sample is therefore probing galaxies near the limit of the NRT \HI\ detectability for SDSS galaxies below log(\Lr) = 8.0.  These galaxies tend to be more distant than galaxies of similar luminosity in other surveys without the NIBLES minimum velocity restriction.  Consequently, our detected galaxies in this low luminosity range tend to be biased toward higher \HI\ mass-to-light ratios (referred to as ``gas-to-light'' ratios henceforth).  In fact, Fig. \ref{fig:lumbins_r_np} shows that for luminosities below log(\Lr) = 8.0, the peak in volume density occurs at \HI\ masses corresponding to a log(\MHI/\Lr) ratio of greater than zero.  For example, the log(\Lr) = 7.75 bin has a peak \HI\ mass volume density in the log(\MHI) = 8.25 bin.  
    
\begin{figure}  
\centering
\includegraphics[width=9cm]{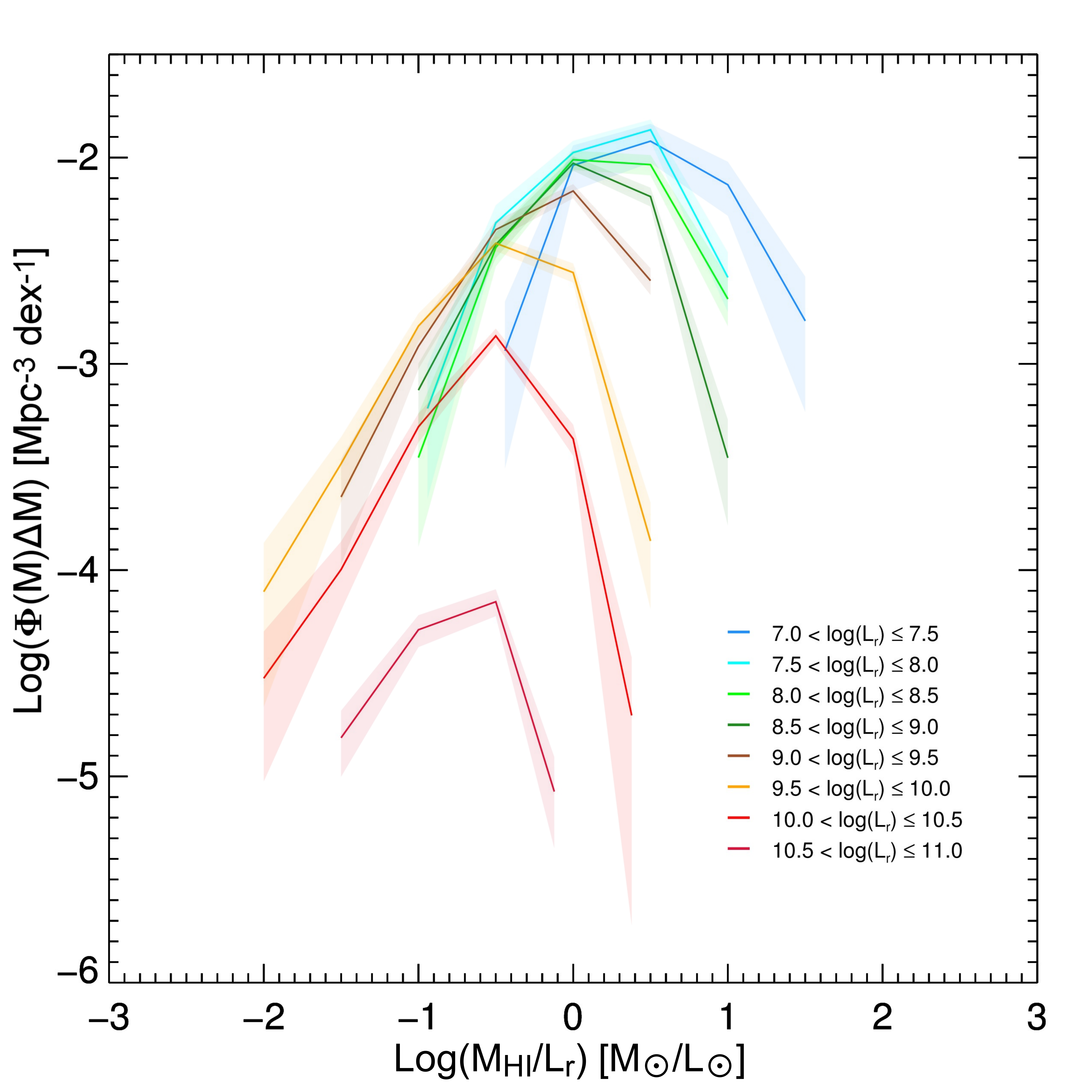}
\caption{\small  log($\Phi$(M)$\Delta$M) in units of Mpc$^{-3}$ dex$^{-1}$ as a function of log(\MHI/\Lr) for the same luminosity bins shown in Fig. \ref{fig:lumbins_r_np}.}
\label{fig:ML_ratio}
\end{figure}
    
This can be more clearly seen in Fig. \ref{fig:ML_ratio} where we show the densities in Fig. \ref{fig:lumbins_r_np} in the form of log(\MHI/\Lr) curves for the same luminosity bins as shown in Fig. \ref{fig:lumbins_r_np}.  We have again excluded the lowest luminosity bin due to poor sampling.  The Luminosity bins below log(\Lr) = 9.0 show a trend of increasing slope with decreasing luminosity for bins above log(\MHI/\Lr) = 0.  This increasing slope indicates that as luminosity decreases, the detections are on average probing more \HI-rich subsamples within each luminosity bin (as to be expected).  The more \HI-poor subsamples, on the other hand, remain undetected, which lowers the measured \HI\ density of the corresponding \HI\ mass bins.  For the low-luminosity bins log(\Lr) = 7.25 - 9.25, this manifests itself as an increased slope for densities below log(\MHI/\Lr) = 0.  The relatively low gas-to-light ratios of these galaxies imply that the densities may be suppressed due to the galaxies within these bins being undetected simply due to distance effects.  Accounting for these non-detections would tend to increase the volume densities for the gas-to-light ratios below log(\MHI/\Lr) = 0, thus increasing the observed slope of the low-mass end of the HIMF.  We explore this further in Sect. \ref{sec:gas-to-light}.  
    
An interesting aspect of these trends in gas-to-light ratio is that for luminosity bins between log(\Lr) = 7.25 and 9.75, the volume density of galaxies increases as a function of gas-to-light ratio at the same rate regardless of luminosity.  In other words, the slopes of the densities as a function of gas-to-light ratio curves are all very similar up to their respective maximum ratios, which depend on luminosity.  Further, the slopes for ratios above the point of maximum density also appear to be relatively similar.
    
\begin{figure}   
\centering
\includegraphics[width=9cm]{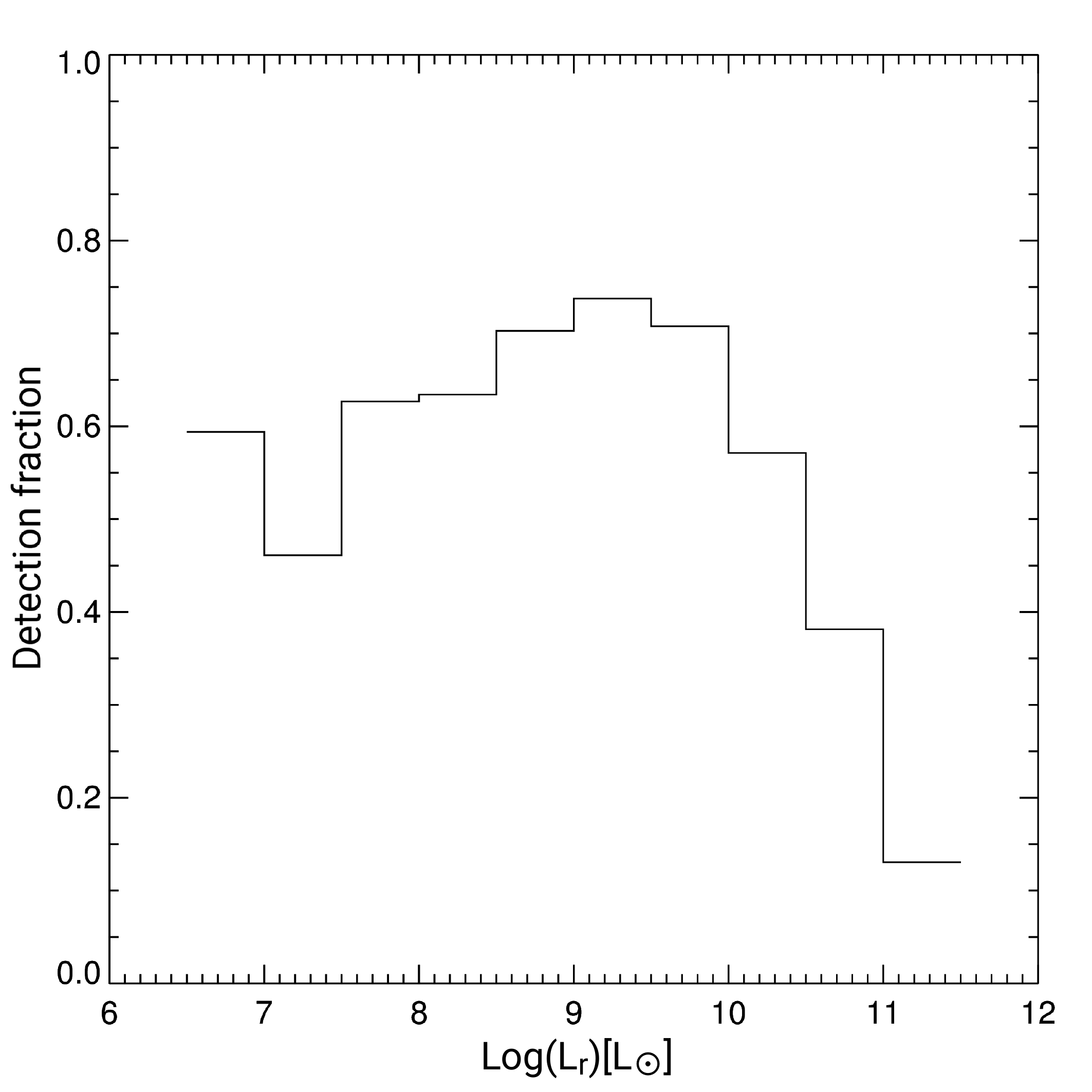}
\caption{\small Detection fraction as a function of $r$-band luminosity.}
\label{fig:NRTdetfrac}
\end{figure}

The increasing gas-to-light ratio with decreasing luminosity trend also manifests itself as a difference in detection fraction ($\sim$10\%) between the log(\Lr) = 9.25 and the lowest luminosity bins as shown in Fig. \ref{fig:NRTdetfrac}.  This is mainly due to the combination of distance effects and sensitivity combined with the NIBLES selection criteria discussed previously.  These low-luminosity galaxies are near the detection limit of both the SDSS and NRT, therefore limiting the both number of sources in the NIBLES sample and the relative number of \HI\ detections.  This decrease in detection fraction is to be expected and is explored further in Appendix \ref{app:detfrac}.  The steeply decreasing detection fraction with increasing luminosity above log(\Lr) = 9.25 (see Fig. \ref{fig:NRTdetfrac}) is mainly due to the increasing \HI\ gas deficiency of larger galaxies.
    
\subsection{Missing galaxies}  
\label{sec:missing_gal}
The second cause of the HIMF faint-end flattening is simply due to the lack of low luminosity galaxies in the sample.  From Fig. \ref{fig:lumbins_r_np} it is evident that any given \HI\ mass bin has the largest contributions from two adjacent luminosity bins, with smaller contributions from additional luminosity bins.  For example, the log(\MHI) = 8.75 bin has equal density contributions from galaxies in the log(\Lr) = 8.25 and 8.75 bins.  However, as we approach the low mass end of the HIMF, these contributions to the \HI\ mass density from adjacent luminosity bins run out.  If we exclude the poorly sampled log(\Lr) = 6.75 bin, the HIMF's slope becomes positive at the low mass end, tracing the rapidly decreasing volume density of the log(\Lr) = 7.25 bin.   
    
As a result of the bias toward high gas-to-light ratio galaxies combined with the lack of available low luminosity sources, we find a flattening of the faint end slope of the HIMF.  However, as we show next, we can make plausible assumptions to correct for the \HI\ mass distributions of the undetected galaxies, as well as the \HI\ mass distributions for galaxies in low-luminosity bins not probed by NIBLES (see Sect. \ref{sec:gas-to-light}) and examine their resulting effects on the HIMF (see \ref{sec:reconstruct}).  To do this, we examine trends in the log(\MHI/\Lr) ratios as a function of luminosity to extrapolate possible distributions in the poorly-sampled and non-sampled low-luminosity bins.
    
\subsection{Gas-to-light ratio distributions}  
\label{sec:gas-to-light}
    
\begin{figure}  
\centering
\includegraphics[width=9cm]{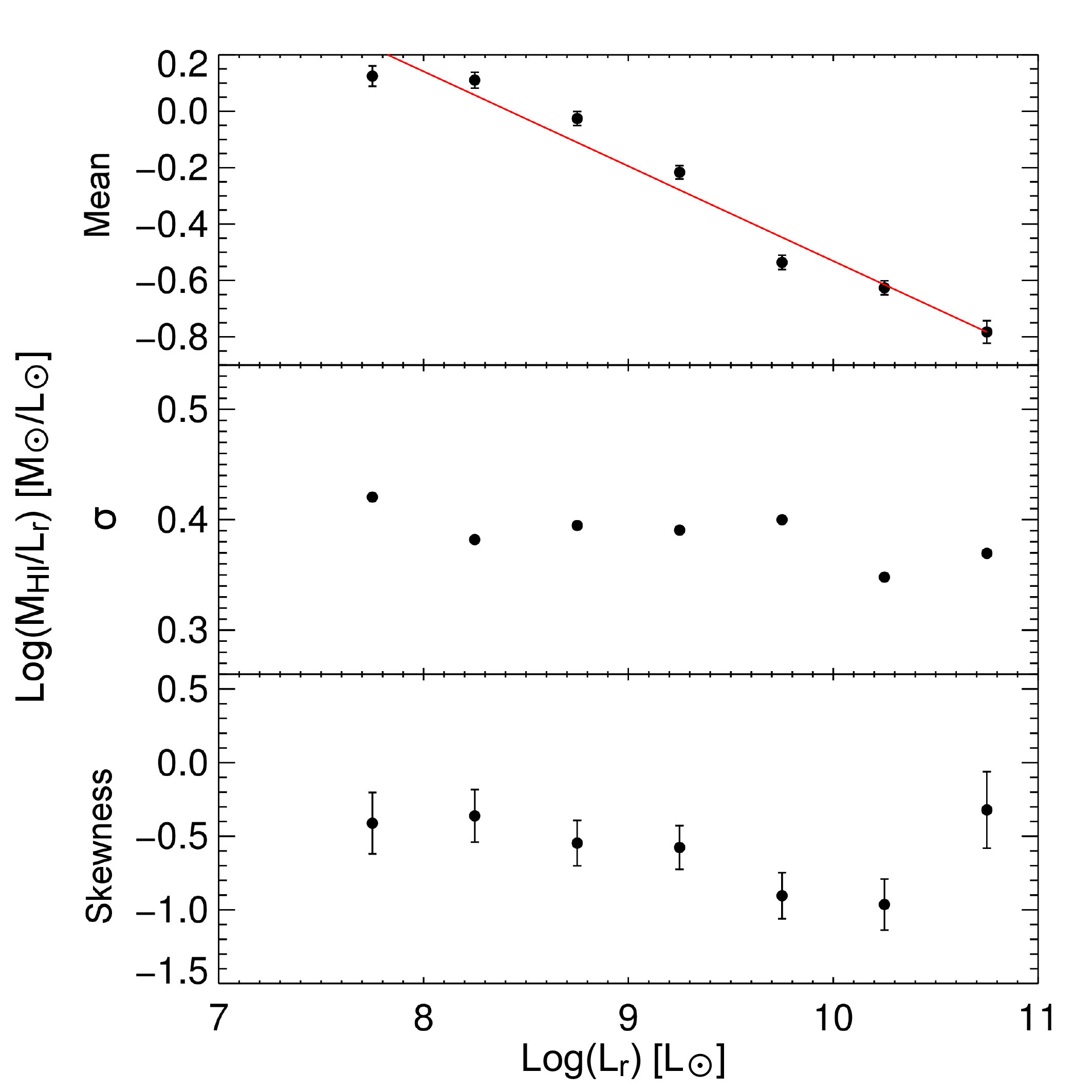}
\caption{\small Top panel: mean log(\MHI/\Lr) ratio for NIBLES detections in each 0.5 dex wide luminosity bin, with uncertainties given by the standard error of the mean.  The straight line indicates the least squares fit to the mean values.  Middle panel: standard deviation ($\sigma$) of the log(\MHI/\Lr) ratio.  Bottom panel:  Skewness of the log(\MHI/\Lr) ratio.  All panels are a function of log(\Lr) in \Lsun.}
\label{fig:Moments}
\end{figure}

The distributions of gas-to-light ratios as a function of luminosity are well constrained in the NIBLES sample.  In Fig. \ref{fig:Moments}, top panel, and also in Fig. \ref{fig:MLratio_Lr} we show that their gas-to-light ratios as a function of $r$-band luminosity follow a fairly consistent trend across the entire luminosity range.  A least-squares fit to the mean log(\MHI/\Lr) values of each 0.5 dex wide luminosity bin gives the relation log(\MHI/\Lr) $= -0.33\ \cdot\ $log(\Lr) + 2.79.
    
While the mean of the gas-to-light ratios in each bin show a trend with luminosity, the same is not true for their standard deviation or skewness (middle and bottom panels of Fig. \ref{fig:Moments}, respectively).  The standard deviation values are all similar (0.39 $\pm$ 0.02 on average) and do not depend on luminosity, while the skewness values display much more scatter (-0.6 $\pm$ 0.3 on average); the given uncertainties are the standard deviations across all the luminosity bins.
    
The skewness values are all rather small, going from about $-0.4$ for bin 7.75 to $-1.0$ for bin 10.25 and could be an indication of a weak trend. We did not consider bins outside the log(\Lr) range 7.75-10.25 range as the 10.75 bin is incompletely sampled and in the bins below 7.75  the galaxies lie only in the upper halves of the bins (see Fig. \ref{fig:Moments}).  The lack of uniformity in these bins affects their measured values for the mean, standard deviation and skewness when compared to the more uniformly sampled bins.  We use the consistent traits across luminosity bins log(\Lr) = 7.75 to 10.25 to construct a reasonable extrapolation of an expected BLF that extends to lower luminosities and compare the results with the HIMFs of Z03 and H11.
    
\subsection{Reconstructed HIMF}  
\label{sec:reconstruct}
    
Due to the detection limits of both the SDSS and NRT as discussed in Sect. \ref{sec:missing_gal}, the NIBLES HIMF displays an approximately flat low-mass slope.  However, since we understand the nature of the data, we can use the observed trend in mean gas-to-light ratio with luminosity together with reasonable estimates of skewness and standard deviation values to create extrapolated gas-to-light distributions in the luminosity bins  that lie outside our sample.
    
The distribution of gas-to-light ratios in each luminosity bin is well described by a Gaussian with negative skewness and an offset corresponding to the slope of the log(\MHI/\Lr) vs. log(\Lr) relation in the previous section, i.e., $\sim$0.3 $\cdot\ $log(\Lr).   As with the discussion in the previous section, Gaussian fits to the log(\MHI/\Lr) ratios from the BLF yield consistent values of variance and skewness across all the luminosity bins from Fig. \ref{fig:Moments}.  We extrapolate these trends down to log(\Lr) = 5.25 due to the increase in average gas-to-light ratio with decreasing luminosity.  Specifically, the density of the log(\MHI)=7.25 bin contains a significant contribution from the log(\Lr) = 5.25 bin.  Additionally, these extrapolated bins are necessary to compensate for the missing low luminosity galaxies as mentioned in Sect. \ref{sec:missing_gal}.
    
We construct the extrapolated gas-to-light distributions such that within each luminosity bin we have:
\begin{equation}
\mathfrak{R}_j = \int_{\Delta M/2L}^{-\Delta M/2L} \phi \\
\textrm{and} \\
\mathfrak{F}_k = \int_{min\ M/L}^{max\ M/L} \phi,
\end{equation}
\noindent where $\mathfrak{R_j}$ is the $j$th gas-to-light ratio bin, $\phi$ represents the skewed Gaussian function, and $\mathfrak{F_k}$ is the detection fraction in the $k$th luminosity bin.  $\Delta M/L$ is the log(\MHI/\Lr) bin size.
    
Uncertainties for the reconstructed bins are calculated from the quadrature sum of the uncertainties in the fit parameters of the skewed Gaussian function.    
    
Once we have reconstructed a gas-to-light distribution for each luminosity bin, we use it to generate the \HI\ mass distribution and recalculate the BLF as outlined in Sect. \ref{sec:method}.
    
In the NIBLES sample, the detection fraction increases with luminosity up to the log(\Lr) = 9.25 bin, which is consistent with sensitivity and distance effects (see Appendix \ref{app:detfrac}).  To analyze the effects of this change in detection fraction with luminosity, we construct two extrapolated BLFs.  The first, ignoring the change in detection fraction with luminosity (our optically corrected sample), and the second taking it into account (our corrected sample).  For the optically corrected BLF, we set the detection fraction of all luminosity bins below 7.75 equal to that of the 7.75 bin, since this is our lowest luminosity, well sampled bin.  We make no attempt to extrapolate a decreasing detection fraction with decreasing luminosity due to the scatter in the observed trend (see Appendix \ref{app:detfrac}).  For the corrected BLF, we set the detection fraction for all bins below 9.25 equal to the detection fraction of the 9.25 bin because the observed fall-off in detection fractions below this bin are consistent with decreases caused by sensitivity and distance effects.
    
\begin{figure}  
\centering
\includegraphics[width=9cm]{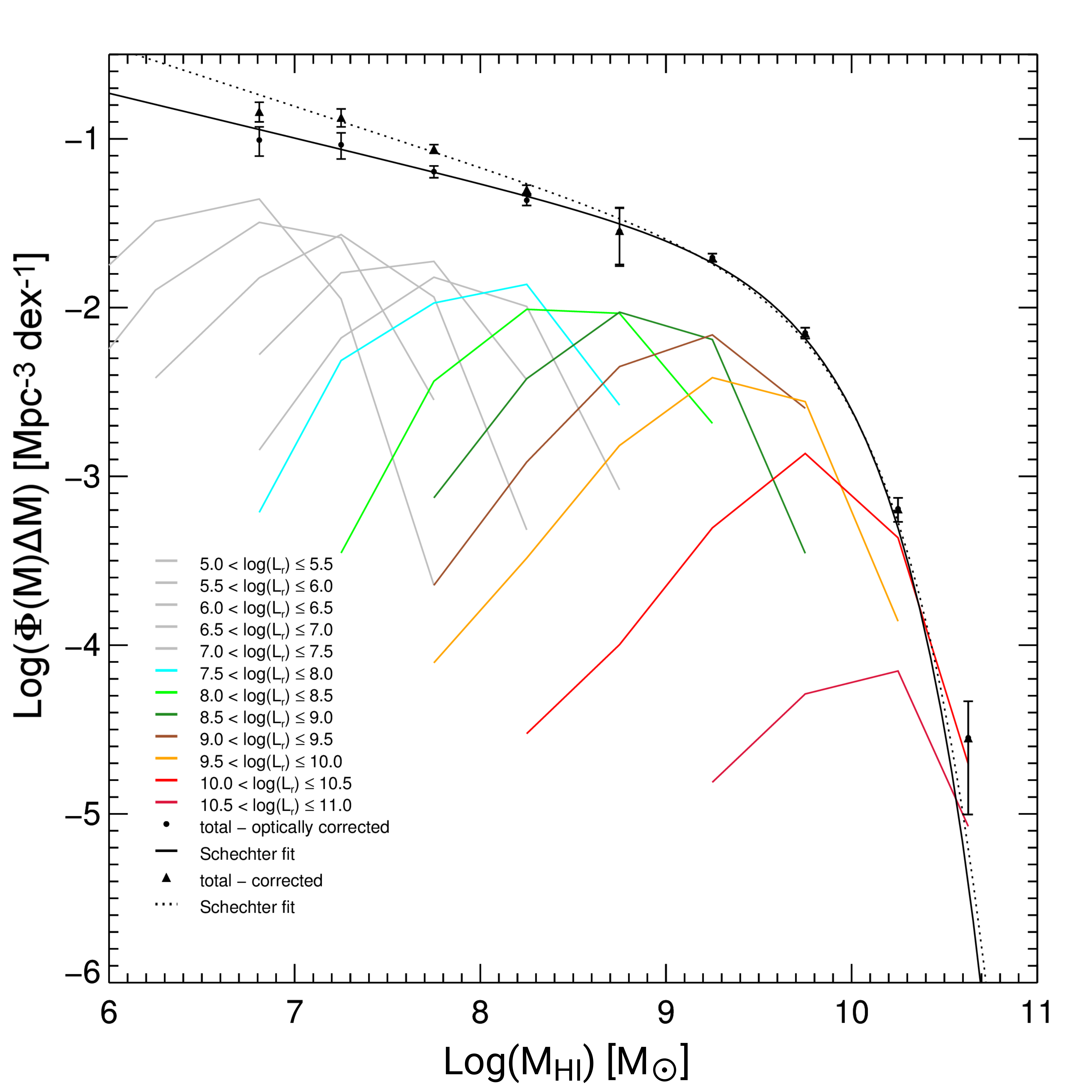}
\caption{\small Reconstructed HIMF per luminosity bin, log($\Phi$(M)$\Delta$M) in units of Mpc$^{-3}$ dex$^{-1}$, as a function of \HI\ mass in each luminosity bin, shown as a set of colored and gray curves.  We recreated a plot similar to Fig. \ref{fig:lumbins_r_np} with artificial \HI\ mass distributions for luminosity bins below log(\Lr) = 7.75 shown in gray.  We omit the uncertainty regions on each luminosity bin for viewing clarity.  Uncertainties for the colored bins are the same as those in Fig. \ref{fig:lumbins_r_np} and uncertainties for the gray bins are given in Table \ref{tab:BLF_corrected}.  Two reconstructed HIMFs are shown: the one based on the optically corrected BLF (which ignores the change in detection fraction with luminosity) is denoted by dots with a solid Schechter fit line, the other based on the corrected BLF (which takes this change into account) is denoted by triangles with a dotted Schechter fit line.}
\label{fig:lumbins_r_fake}
\end{figure}

In Fig. \ref{fig:lumbins_r_fake} we show (in gray) the reconstructed HIMFs per luminosity bin derived from the optically corrected BLF, together with the reconstructed HIMF from the corrected BLF for comparison.  The shaded uncertainty regions from Fig. \ref{fig:lumbins_r_np} are omitted here for clarity.  The low-luminosity sources shown here in gray have the effect of boosting the low-mass slope of the HIMF to a level agreeing with the observed slopes in recent blind \HI\ surveys.  However, even using the extrapolated bins down to log(\Lr) = 5.25, there is an apparent turn-over beginning at the log(\MHI) = 7.25 bin due to the lack of sources at even lower luminosity.  This is illustrated by the log(\MHI) = 7.25 point falling below the HIMF Schechter fit in Fig. \ref{fig:lumbins_r_fake}.  These extrapolations show the need to probe extremely low luminosity sources when attempting to construct an HIMF from optically selected sources.  
    
The need to probe extremely low luminosity galaxies is not unique to our survey: it applies to all surveys used to derive an HIMF. However, as blind \HI\ surveys are not attempting to recover a two-dimensional \HI-optical BLF distribution using an \HI\ survey of optically selected galaxies like NIBLES, their lack of data at the low luminosity end is not problematic in this regard.
    
The difference in low-mass slope between the optically corrected and corrected BLF HIMFs is evident.  The approximately 10\% difference in detection fraction between the log(\Lr) = 9.25 and 6.75 bins manifests itself as an approximately 6\% steeper low-mass slope in the HIMF.  The Schechter fit parameters for the two HIMFs are: \\
\indent optically corrected: $\Phi = 0.0085 \pm 0.0015$, log($M_{\star}$) = $9.72 \pm 0.06$, $\alpha = -1.26 \pm 0.04$ \\
\indent corrected: $\Phi = 0.0066 \pm 0.0012$, log($M_{\star}$) = $9.79 \pm 0.07$, $\alpha = -1.37 \pm 0.03$ 
    
These results imply that the fall-off in detection fraction for luminosity bins below log(\Lr) = 9.25 is due primarily to sensitivity and distance effects.
    
For our corrected BLF, volume density values for luminosity bins of 8.25 and below are listed in Table \ref{tab:BLF_corrected} as log($\Phi$(\MHI,\Lr)$\Delta M$ $\Delta L$) in units of Mpc$^{-3}$ dex$^{-1}$, together with their fractional uncertainties.
\onecolumn
\begin{table}
\centering
\caption{$r$-band luminosity-\HI\ mass distribution function for log(\Lr) bins $\leq$8.25, for the corrected BLF}
\small
\renewcommand{\tabcolsep}{1pt}  
\begin{tabular}{ |l||c|c|c|c|c|c|c|}
    \hline
    & \multicolumn{7}{|c|}{log(\Lr) bin [\Lsun]} \\
    \hline
    log(\MHI) bin [\Msun] & 5.25 & 5.75 & 6.25 & 6.75 & 7.25 & 7.75 & 8.25 \\
    \hline
    \hline
    5.25&-2.64 $\pm$ 1.36&-3.39 $\pm$ 2.72&&&&& \\
    \hline
    5.75&-1.84 $\pm$ 0.38&-2.42 $\pm$ 0.87&-3.11 $\pm$ 1.81&&&& \\
    \hline
    6.25&-1.32 $\pm$ 0.08&-1.73 $\pm$ 0.22&-2.25 $\pm$ 0.55&-2.87 $\pm$ 1.19&-3.61 $\pm$ 2.41&&-5.14 $\pm$ 2.22 \\
    \hline
    6.75&-1.19 $\pm$ 0.04&-1.32 $\pm$ 0.05&-1.65 $\pm$ 0.12&-2.11 $\pm$ 0.33&-2.67 $\pm$ 0.76&-3.21 $\pm$ 0.27&-5.08 $\pm$ 2.11 \\
    \hline
    7.25&-1.78 $\pm$ 0.15&-1.42 $\pm$ 0.03&-1.40 $\pm$ 0.04&-1.62 $\pm$ 0.06&-2.01 $\pm$ 0.18&-2.31 $\pm$ 0.09&-3.45 $\pm$ 0.27 \\
    \hline
    7.75&-3.47 $\pm$ 0.67&-2.38 $\pm$ 0.30&-1.77 $\pm$ 0.08&-1.56 $\pm$ 0.03&-1.65 $\pm$ 0.04&-1.97 $\pm$ 0.06&-2.44 $\pm$ 0.08 \\
    \hline
    8.25&&&-3.15 $\pm$ 0.50&-2.26 $\pm$ 0.20&-1.82 $\pm$ 0.04&-1.86 $\pm$ 0.05&-2.01 $\pm$ 0.05 \\
    \hline
    8.75&&&&&-2.91 $\pm$ 0.36&-2.58 $\pm$ 0.14&-2.03 $\pm$ 0.05 \\
    \hline
    9.25&&&&&&&-2.69 $\pm$ 0.11 \\
    \hline
    9.75&&&&&&&-5.31 $\pm$ 2.28 \\
    \hline
\end{tabular}

\label{tab:BLF_corrected}

Volume density values for the corrected BLF, for the luminosity bins at and below log(\Lr) = 8.25.  The listed uncertainties are fractional.
\end{table}
\twocolumn

\section{Discussion}  
\label{sec:discussion}
In addition to being useful tools for measuring the baryon density of the local universe, both the HIMF and optical LF have also been instrumental in providing constraints on galaxy formation models \citep[see, e.g.,][]{Lagos2011,duffy2012,dave2013}.  Currently these constraints are used independently, but our bivariate \MHI-\Lr\ distributions provide cross-constraints on these models that will provide additional insights into the physical processes necessary to realize these properties.  Additionally, the small variation in the observed standard deviation and skewness values of the gas-to-light ratios in each luminosity bin can provide additional modeling constraints.
    
As ours is the first optical luminosity-\HI\ mass bivariate distribution constructed, we have run checks on its consistency through comparisons with published HIMFs derived in other studies.  Because each of these studies has its own design, there are a number of factors that need to be accounted for when making these comparisons.  We begin by comparing our HIMFs with those of blind \HI\ surveys in Sect. \ref{sec:blind_surveys}, followed by a comparison between expected gas-to-light ratio distributions from our BLF and recently studied almost dark galaxies in Sect. \ref{sec:almost_dark}.  We compare our uncorrected HIMF with others from optically selected samples in Sect. \ref{sec:opt_surveys}, followed by a discussion and comparison of our $\Omega_\textrm{HI}$ with those from other surveys.

\subsection{Comparison with HIMFs from blind \HI\ surveys}  
\label{sec:blind_surveys}
    
\begin{figure}  
\centering
\includegraphics[width=9cm]{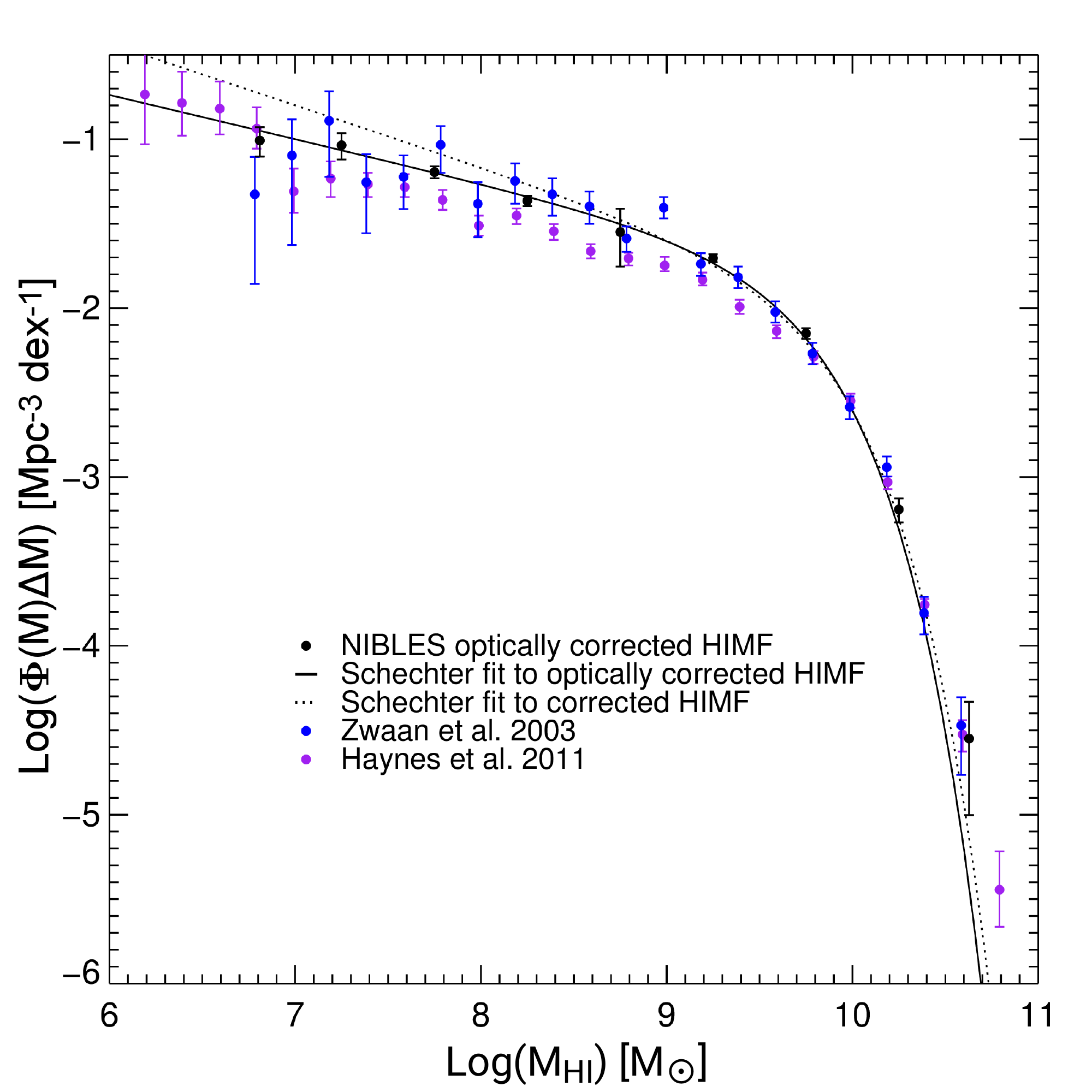}
\caption{\small Comparison of the NIBLES optically corrected HIMF (black dots denote the data points, and the black line the Schechter fit) and the Z03 and H11 blind \HI\ survey HIMFs (blue and purple dots denote the data points from Z03 and H11 respectively, which have been rescaled to match the NIBLES flux scale (see text) and Hubble constant used in this paper).}
\label{fig:zwaan_haynes_nib}
\end{figure}

Blind \HI\ survey HIMFs have become one of the standard litmus tests used in comparisons of cosmological simulations with observations.  Here (see Fig. \ref{fig:zwaan_haynes_nib}) we compare our results from the optically corrected BLF to those of Z03 and H11. 

Three differences between the surveys have either negligible effects or require simple adjustments.  First, although H11 uses peculiar velocity corrections to the Hubble flow, this difference with our method has a negligible impact on the comparison since the largest discrepancies will occur for recession velocities below the minimum NIBLES velocity of 900 \kms.  Second, we have applied offsets to the Z03 and H11 HIMFs (+0.07 and +0.11 dex in log(\MHI), respectively) corresponding to the flux scale differences between those surveys and NIBLES to bring them into agreement with the \citet{oneil04b}  flux standard -- see \citetalias{vandriel2016} for details.  Third, we have rescaled the Z03 data to the 70 \kmsMpc\ Hubble constant adopted in this paper and \citetalias{vandriel2016}.
    
There is good agreement in Fig. \ref{fig:zwaan_haynes_nib} between our optically corrected HIMF and those of Z03 and H11.  Although the low-mass data (log(\MHI) = 7.5 - 9.5) show small systematic vertical offsets, the slope of our optically corrected HIMF is in very good agreement with those of the other two, only differing by about 0.01 (see also Sect. \ref{sec:reconstruct}).  The agreement between the HIMFs is brought about by our re-scaling of the Z03 and H11 data flux scales.

Despite the suggestion in \cite{martin2010} and H11 that ALFALFA found a 1 dex greater volume density of high-mass sources than the HIPASS survey, after adjusting both to the same calibration standard, the high-mass data points appear to be in good agreement as shown in Fig. \ref{fig:zwaan_haynes_nib}.  
 \HI\ surveys have become sufficiently large with statistically significant numbers of rare high-mass galaxies that calibration differences have a significant impact on their interpretation.
    
The NIBLES optically corrected HIMF was constructed from trends observed in the gas-to-light ratio distributions, utilizing an extrapolation for galaxies that lie outside the observable limits (see  Sect. \ref{sec:gas-to-light}) of the NIBLES sample.  This extrapolation appears justified by the fact that the reconstructed low-mass volume density agrees well with those of the Z03 and H11 blind surveys.  To achieve this agreement we included extrapolations down to the log(\Lr) =  5.25 bin (see Fig. \ref{fig:lumbins_r_np}). This corresponds to an $r$-band absolute magnitude of --8.5, i.e., in the same range as globular clusters.  The need to extrapolate to such low-luminosity galaxies indicates that optically selected surveys would have to achieve impractically high sensitivity to accurately sample the low-mass end of the HIMF, as will be discussed further in Sect. \ref{sec:opt_surveys}.

We discussed in Sect. \ref{sec:missing_gal} how the sensitivity limits prohibiting detection of very low luminosity objects suppresses the low-mass slope of the HIMF.  This is true even for the optically corrected BLF, which agrees well with the HIMFs of Z03 and H11.  After fully correcting for the sensitivity effects discussed in Sects. \ref{sec:HI_sens_lim} and Appendix \ref{app:detfrac}, the resulting corrected BLF produces an HIMF with an approximately 6\% steeper low \HI-mass slope, in agreement with that found in \citet{zwaan2005}. This steeper slope reflects the contribution of some extremely low luminosity \HI-rich objects, examples of which have been encountered in several \HI\ surveys.

\subsection{Almost dark galaxies}  
\label{sec:almost_dark}
Blind \HI\ surveys have detected a number of low-luminosity but \HI-rich galaxies, some of which initially appeared to be optically dark sources.  However, the \HI\ and optical properties of these sources appear to be consistent both with the gas-to-light ratios predicted by our BLF, as well as the upper-most limits of \MHI/L found in other surveys \citep[e.g.,][]{rosenberg2005, doyle2005}.

As noted before, the \HI\ detectability of ALFALFA is comparable to that of NIBLES, but as a blind \HI\ survey ALFALFA can be used to search for extragalactic sources without optically cataloged counterparts, and it can detect very nearby objects since it has no lower radial velocity limit.
Follow-up studies in both the optical and with radio interferometers have been made of ALFALFA detections that initially appeared to be (nearly) without optical counterparts. These comprise ALFALFA (Almost) Dark Galaxies Project candidates including SHIELDS (Survey of \HI\ in Extremely Low-mass Dwarfs) objects, and the Leo P dwarf.  Deep optical images usually revealed a likely optical counterpart, but not always \citep[see, e.g.,][]{Cannon2011, Giovanelli2013, Rhode2013, Cannon2015, Janowiecki2015, McNichols2016}.

Among the ALFALFA (Almost) Dark Galaxies pilot surveys, we compare the gas-to-light ratios of the two isolated objects from \citet{Cannon2015} and the sole optically detected galaxy from \citet{Janowiecki2015} with those from the corrected BLF.  In our BLF,  both of the \citet{Cannon2015} galaxies, with luminosities in the log(\Lr) = 7.25 and 6.75 bins, fall in the log(\MHI/\Lr) = 1 bin.  According to our two-dimensional corrected BLF distribution, we would expect to find $\sim$30 to 40\% of \HI\ detected galaxies of similar luminosities in this gas-to-light ratio bin.  The optical detection from \citet{Janowiecki2015}  (in the log(\Lr) = 7.25 and log(\MHI/\Lr) = 1.5 bins), is where we would expect to find only $\sim$2\% of galaxies of similar luminosity.   However, it should be noted that the \MHI/\Lr\ ratio of the latter source is likely to be overestimated, as it clearly contains optical flux outside the measurement apertures used in the paper.

Correlating the $\alpha$.40 catalog to optical sources from the SDSS DR9 give results that are in reasonable qualitative agreement with our BLF.  We only used sources with $r$-band petrosian radii larger than 3’’, as sources with smaller radii tend to have unreliable magnitudes.  This optical correlation results in $\sim$ 25\% and 35\% of galaxies in the log(\Lr) = 6.75 and 7.25 bins respectively being contained in the log(\MHI/\Lr) = 1 bin.  For the log(\Lr) = 7.25 bin, $\sim$ 6\% of the galaxies are in the log(\MHI/\Lr) = 1.5 bin.  

An order of magnitude smaller gas-to-light ratio (log(\MHI/\Lr) = 0.2) was found for the nearby Leo P dwarf, at a distance of only 1.6 Mpc.  With a log($L_{\rm V}$)$\sim$5.8 it is one of the faintest optical ALFALFA detections.  It also has one of the lowest \HI\ masses observed at log(\MHI)$\sim$6 \citep[see, e.g.,][]{Giovanelli2013, Rhode2013, Bernstein2014, McQuinn2015}.  The gas-to-light ratio of the Leo P dwarf is also consistent with our corrected BLF.  Based on our extrapolations to such low luminosities, roughly 27\% of the population of these sources would have gas-to-light ratios in the log(\MHI/\Lr) = 0 bin or lower.  

From the comparison with our BLF, it appears that the four (Almost) Dark Sources discussed above have fairly common gas-to-light ratios for their luminosities.  The reason they are (almost) dark is simply that at such low luminosities the mean gas-to-light ratio grows quite large, making these sources relatively easier to detect in \HI\ surveys.

To illustrate this point, a galaxy with a high gas-to-light ratio (log(\MHI/\Lr) $\sim$ 1.5) with log(\Lr) $\sim$ 5 would be detectable to only about 2 Mpc in the SDSS but out to 13 Mpc in \HI. This comparison is based on the assumption of \Wfifty=36 \kms, an rms of 2.3 mJy and a signal to noise ratio of three, and an estimated SDSS $r$-band galaxy detection limit of 17.77 mag (see  \citep{loveday2002}); for NIBLES, we also noted that galaxies fainter than this limit often have unreliable magnitudes (see \citetalias{vandriel2016} for details).

These differences in detectability in optical vs. \HI\ for these high gas-to-light ratio galaxies explains the absence of an optical counterpart in \HI\ surveys.  While these galaxies make interesting case studies for galaxy formation and evolution, they so far do not appear to contain exceptionally high gas-to-light ratios given their low luminosities.

\subsection{Comparison with HIMFs from optical samples}  
\label{sec:opt_surveys}
NIBLES is the first \HI\ survey to use optically selected sources that were subsequently followed up with a uniform set of \HI\ observations.  This is in contrast to previous works such as R93 and S05 which used a combination of archival optical and \HI\ data.  
    
R93 used published conversions from optical luminosities to \HI\ masses for different classes of galaxies to derive an HIMF from existing literature sources, while S05 used 2771 spiral and irregular galaxies from the Uppsala General Catalog of Galaxies \citep[UGC: ][]{UGC} with optical diameters greater than 1$'$.  The closest equivalent in our study is the uncorrected HIMF, which does not attempt to correct for the missing low luminosity sources.
    
\begin{figure}  
\centering
\includegraphics[width=9cm]{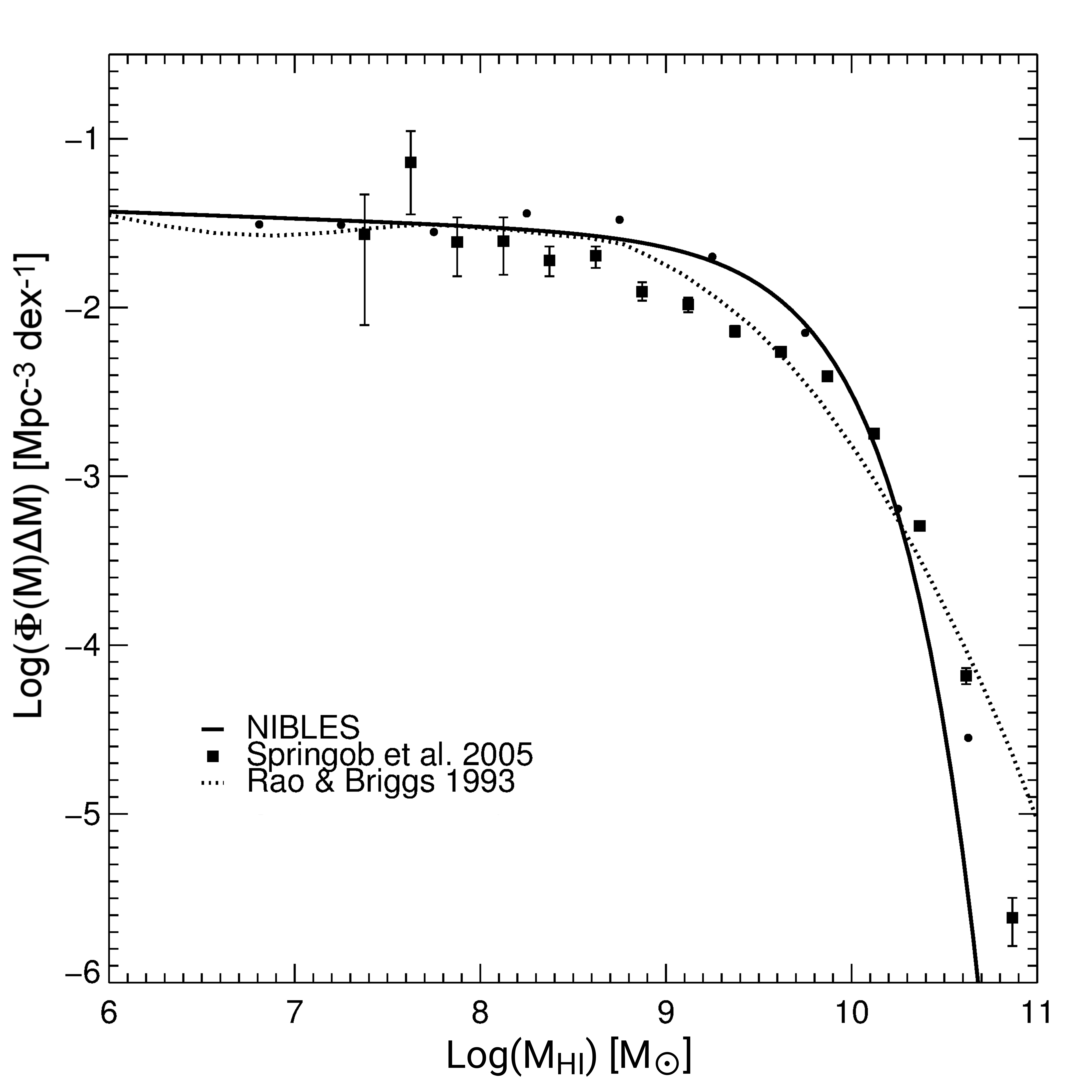}
\caption{\small NIBLES uncorrected HIMF from Fig. \ref{fig:lumbins_r_np} (solid line) compared to the HIMFs derived from optical samples by S05 (squares, with error bars) and R93 (dotted line).  See the text for further details.}
\label{fig:rao}
\end{figure}

The HIMFs obtained by these differing approaches can be seen in Fig. \ref{fig:rao}.  The S05 data correspond to the $\sum 1/V_{max}$ method results in that paper, and for the R93 fit we converted the HIMF from their Fig. 5 from \HI\ mass density to the now commonly used number density and scaled their data to the Hubble constant of 70 \kmsMpc\ used in NIBLES.
    
At low \HI-masses (log(\MHI)$<$8.5) the density values are relatively similar for all three studies. However, near the knee of the Schechter functions, above log(\MHI) $\sim$ 8.5, a discrepancy appears with the NIBLES uncorrected HIMF having a higher volume density than the other two. Considering that the NIBLES uncorrected HIMF has density values in good agreement with the blind \HI\ surveys, for the high \HI-mass galaxies (log(\MHI) $\sim$ 8.5 - 10), this comparison suggests that the morphology and color criteria of R93 and S05 introduced biases. The NIBLES approach, which did not select by either morphology or color, results in a significantly more complete sampling of the volume density of even luminous, high-\HI-mass galaxies than selecting optical sources generally expected to contain \HI.  

For the low \HI-mass galaxies (log(\MHI) $<$ 8.5) however, the similarities between the low \HI\ mass volume densities in all three studies combined with the analyses presented in Sects. \ref{sec:gas-to-light} and \ref{sec:blind_surveys} suggest that optically selected samples generally lack sufficiently low-luminosity sources to construct an HIMF with the same low-mass slope as derived from blind \HI\ surveys.

\subsection{Cosmic density of \HI, $\Omega_\textrm{HI}$}  

We can use our \HI\ Mass Function to calculate the cosmic mass density of \HI,  $\Omega_{\textrm{\HI}}$.  We find $\Omega_{\textrm{\HI}} =$ 4.2 $\pm$ 0.2  from the corrected HIMFs, in units of $10^{-4} h_{70}^{-1}$. This differs by only 0.1 from the value obtained from our uncorrected HIMF, and it is also very similar to the values obtained from other surveys:
$4.0 \pm 0.5$  derived from HIPASS detections by Z05 and $4.2-4.4 \pm 0.1$  derived from the summation of binned measurements by \cite{martin2010} and H11.
The only measured 21cm \HI\ line $\Omega_{\textrm{\HI}}$ value at a significantly higher redshift, $z \sim 0.32$, is  5.0 $\pm 1.8 \times 10^{-4} h_{70}^{-1}$ \citep{rhee2016} is based on stacked GMRT \HI\ spectra of 165 galaxies with known optical redshifts. So the 21cm $\Omega_{\textrm{\HI}}$ remains constant within the estimated uncertainties out to at least $z \sim 0.3$, but at a level which is only about half of that measured from Damped Lyman Alpha (DLA) systems at $z  >  2$  \citep[see][and
references therein]{rhee18}.

Theoretical studies of the cosmic evolution of $\Omega_{\textrm{\HI}}$ have yielded a range of results. Since our estimate, while agreeing with those from other studies (see above, e.g., Z05), has formally smaller uncertainties, it can thus be used to constrain theoretical studies. For example, the theoretical HIMF of \citet{popping15} over-predicts the number of low \HI\ mass galaxies compared to our results, but this discrepancy is less compared to the HIMFs of \citet{martin2010} or H11. As a consequence, \citet{popping15} over-predicts $\Omega_{\textrm{\HI}}$ by about a factor of four. On the other hand, more recent models, such as those of \citet{dave17}, under-predict $\Omega_{\textrm{\HI}}$ by about 30\% compared to our observed results. The latter difference between an observational and a theoretical estimation is significant at the $\sim$6$\sigma$ level (we have a $\approx$5\% uncertainty). The discrepancy likely arises due to the model of \citeauthor{dave17} preferentially under-predicting the \HI\ gas fraction of high mass galaxies (log(\Mstar) $\ga$ 10.5) compared to observations, like our NIBLES results \citepalias{vandriel2016}. 

Underlying models predicting the \HI\ content can be divided into two basic types of prescriptions -- either pressure-based for regulating the phase relative fractions of atomic to molecular gas \citep{blitz2004},  or dependent on the intensity of the interstellar radiation field molecular cloud opacity \citep{krumholz2009}. While both theoretical approaches explain certain aspects of the gas and star formation properties of galaxies, it is not clear that they provide appropriate descriptions of the integrated properties of the ensemble of galaxies or even, in phenomenological models of galaxy formation and evolution, that their discrepancies with observed HIMFs and $\Omega_{\textrm{\HI}}$ values are mainly driven by problems with these theoretical prescriptions. 

One thing is certain, determining a robust and accurate characterization of the local \HI\ mass function and $\Omega_{\textrm{\HI}}$ through the BLF, as we have done, does provide additional needed constraints on the gas cycle in galaxies -- how gas is acquired, what processes regulate the gas phases within galaxies, and how gas is lost -- and thus on the physics of galaxy evolution.

\section{Future work}  
\label{sec:future}
    
The Bivariate Luminosity Functions presented in this work are based on \HI\ detection fractions of optically selected sources.  These detection fractions drop off as a function of decreasing gas-to-light ratio, as one would expect.  However, there is a currently unknown \HI\ distribution lying below the detection threshold of NIBLES which will expand the plane of the BLF if probed.  We are currently working on implementing a BLF incorporating four times more sensitive follow-up observations of the NIBLES undetected sample with the Arecibo L-band wideband receiver \citep[see also the pilot survey results in][]{butcher2016}.
    
Further investigation of the BLF presented here is possible using the ALFALFA survey data, even though it has about the same \HI\ sensitivity as NIBLES.  The low luminosity and low \HI\ mass portion of the BLF plane could be explored using the large statistics provided by the ALFALFA sources.  However, the optical properties of many of these low luminosity sources are not accurately measured in SDSS and would require the re-processing of all photometric sources with $r$-band Petrosian magnitudes greater than 17.77.  Additionally, follow-up observations (utilizing the L-band wideband receiver) of low luminosity sources not detected in ALFALFA would also be required to accurately sample the BLF plane in this parameter space.  

The NIBLES BLF can also be used to construct a trivariate luminosity--\HI-mass--dynamical mass function of galaxies utilizing line widths to infer the dynamical mass.  This trivariate function could be used to examine what role the dark matter halo plays in regulating galaxy evolution.

\section{Conclusions}  
\label{sec:conclusion}
We have developed the first optical luminosity-\HI\ mass BLF  based on the NIBLES sample of SDSS galaxies in the local volume (900 \kms < $cz$ < 12 000 \kms), selected to span their entire range of stellar masses.  The HIMF constructed from our uncorrected BLF, which was derived using only \HI\ detections, agrees very well with other HIMFs constructed from optically selected samples.
    
We observed a very consistent gas-to-light ratio distribution (with similar standard deviations and skewness) displaying a progressive increase in the mean log(\MHI/\Lr) value with decreasing luminosity across the entire luminosity range of the NIBLES sample, which may suggest that galaxy populations in all luminosity bins undergo the same evolutionary processes. 
    
Using the predictable offset in mean gas-to-light ratio and the consistency in gas-to-light distribution as a function of luminosity, we constructed an optically corrected BLF by extrapolating these properties in luminosity bins down to log(\Lr) = 5.25.  Estimating bivariate densities down to this low luminosity results in a corrected HIMF with the same low mass slope as those derived from the blind \HI\ surveys Z03 and H11.  This suggests that the low mass end of the HIMF may contain significant contributions from galaxies with very low luminosities.
    
We also observed a $\sim$10\% decline in detection fraction from log(\Lr) = 9.25 to the lowest luminosity sources  due to distance effects and sensitivity.  Correcting  for this decline leads to a slightly more negative low mass slope of the HIMF, in agreement with the \cite{zwaan2005} HIPASS results.  

The combination of sensitivity and minimum radial velocity cut-offs makes the detection of very low \HI\ mass sources difficult for both the NIBLES and HIPASS surveys. This is reflected in the \cite{zwaan2005} HIPASS HIMF low mass slope, devoid of sources below log(\MHI) = 7. The NIBLES corrected HIMF, on the other hand, has a low mass slope obtained by accounting for the above mentioned trends in our data as well as observational limitations.  We are therefore able to improve on HIMFs from previous optically selected \HI\ surveys and produce HIMFs that are in agreement with those of other blind \HI\ surveys.  

We note that based on the differences in detectability in the optical vs. \HI\ for relatively high gas-to-light ratio sources with low luminosity, the optical counterparts of such \HI\ detections may be very difficult to detect.  Indeed, this appears to be the case with the ALFALFA (almost) Dark Galaxies observed to date.
    
Our corrected BLF and resulting HIMF explain why HIMFs tend to have steeper low mass slopes than the faint end of optical LFs: the increasing mean gas-to-light ratios of the lower luminosity sources cause a source's \HI\ mass to be counted in higher \HI\ mass bins relative to where it would be counted in an optical luminosity function.  This increases the density relative to luminosity, resulting in a boosted low-mass slope of the HIMF.  
    
Optically selected samples like NIBLES  appear to lack a significant percentage of low luminosity objects that are detectable in \HI, whereas our optically corrected and corrected BLFs can reproduce blind \HI\ survey HIMFs, and our $\Omega_{\textrm{\HI}}$ agrees with other surveys.  All this gives us confidence that our corrected BLF provides an accurate representation of the galaxy population in the local universe.  Therefore, it can be used to provide further constraints on the gas cycle in galaxies and insights into galaxy evolution.  

Investigation of the low luminosity (log(\Lr) $<$ 7) regime of the BLF with higher sensitivity optical surveys could further confirm the accuracy of our extrapolated gas-to-light ratio distributions.
 
\begin{acknowledgements}
The \nan\ Radio Telescope is operated as part of the Paris Observatory, in association with 
the Centre National de la Recherche Scientifique (CNRS) and partially supported by the 
R\'egion Centre in France.
	
Funding for SDSS-III has been provided by the Alfred P. Sloan Foundation, the Participating Institutions, the National Science Foundation, and the U.S. Department of Energy Office of Science. The SDSS-III web site is http://www.sdss3.org/.
	
SDSS-III is managed by the Astrophysical Research Consortium for the Participating Institutions of the SDSS-III Collaboration including the University of Arizona, the Brazilian Participation Group, Brookhaven National Laboratory, Carnegie Mellon University, University of Florida, the French Participation Group, the German Participation Group, Harvard University, the Instituto de Astrofisica de Canarias, the Michigan State/Notre Dame/JINA Participation Group, Johns Hopkins University, Lawrence Berkeley National Laboratory, Max Planck Institute for Astrophysics, Max Planck Institute for Extraterrestrial Physics, New Mexico State University, New York University, Ohio State University, Pennsylvania State University, University of Portsmouth, Princeton University, the Spanish Participation Group, University of Tokyo, University of Utah, Vanderbilt University, University of Virginia, University of Washington, and Yale University. 
	
We thank the anonymous referee for their helpful comments and suggestions for improving this paper.	
\end{acknowledgements}    
    
\bibstyle{aa}
\bibliographystyle{aa}
    
\bibliography{NIBLES_bivar}
    
\begin{appendix}

\section{NIBLES detection fractions}  
\label{app:detfrac}
    
The fraction of galaxies detected for NIBLES peaks at the log(\Lr) = 9.25 bin and decreases from there with both increasing and decreasing luminosity (see Fig. \ref{fig:NRTdetfrac}).  The decrease in detection fraction with increasing luminosity is to be expected since more luminous galaxies tend to be relatively gas poor compared to their lower luminosity counterparts.  However, the decrease in detection fraction with decreasing luminosity is somewhat counter-intuitive since the average gas-to-light ratio increases with decreasing luminosity.  However, as mentioned in Sect. \ref{sec:results}, this can be explained by a combination of the NRT sensitivity and the NIBLES selection criteria.  We give a more detailed description of this phenomenon here.  
    
We test the effect of our selection criteria on our detection fraction as follows. 
    
\begin{figure}
\centering
\includegraphics[width=8.5cm]{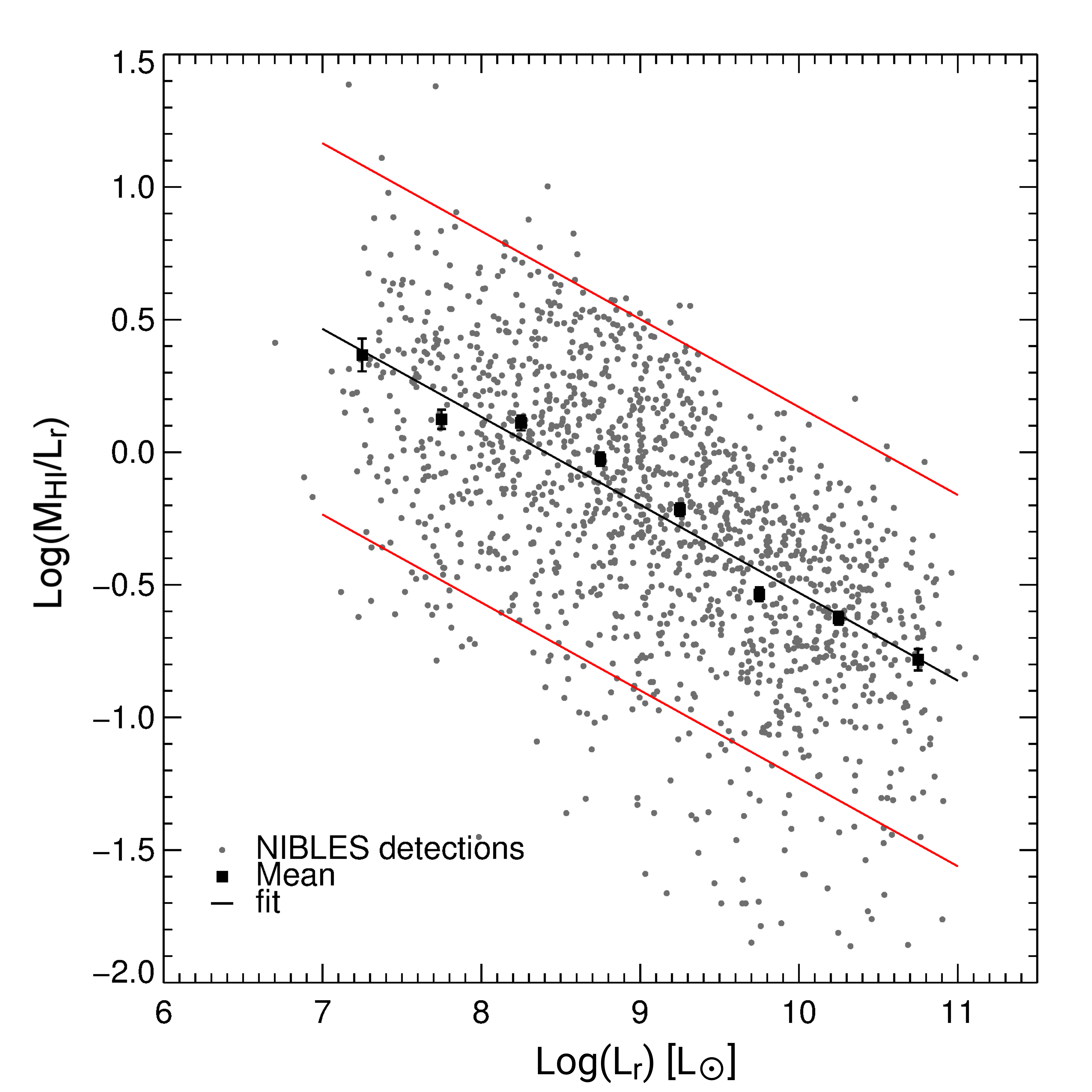}
\caption{\small Log(\MHI/\Lr) ratio as a function of log(\Lr) for the NIBLES sample.  Gray dots indicate individual sources while black squares show the mean ratio in each 0.5 dex wide log(\Lr) bin.  The black line is the least squares fit to the mean ratios while the red lines are the upper and lower bounds encompassing 92\% of the sources, which corresponds to shifting the mean fit by $\pm$ 0.7 dex.}
\label{fig:MLratio_Lr}
\end{figure}

In Fig. \ref{fig:MLratio_Lr} we show the gas-to-light ratio (log(\MHI/\Lr)) as a function of \Lr\ for NIBLES detections.  Our fit to the data is shown in black, along with upper and lower bounds around 92\% of the data in red. For each galaxy, we randomly assign an \HI\ mass based on its luminosity that falls within these two bounds.  \HI\ surveys in general do not have a set minimum detectable \HI\ line flux due to the  dependence of detectability on line width.  NIBLES has the added complication that we do not have a uniform $rms$ noise for each source either.  Therefore, for illustration purposes, we assume a 0.2 \Jykms\ minimum line flux for detectability, which is consistent with the lowest fluxes detected in NIBLES.  This gives us a minimum detectable \HI\ mass as a function of distance.
    
\begin{figure}
\centering
\includegraphics[width=8.5cm]{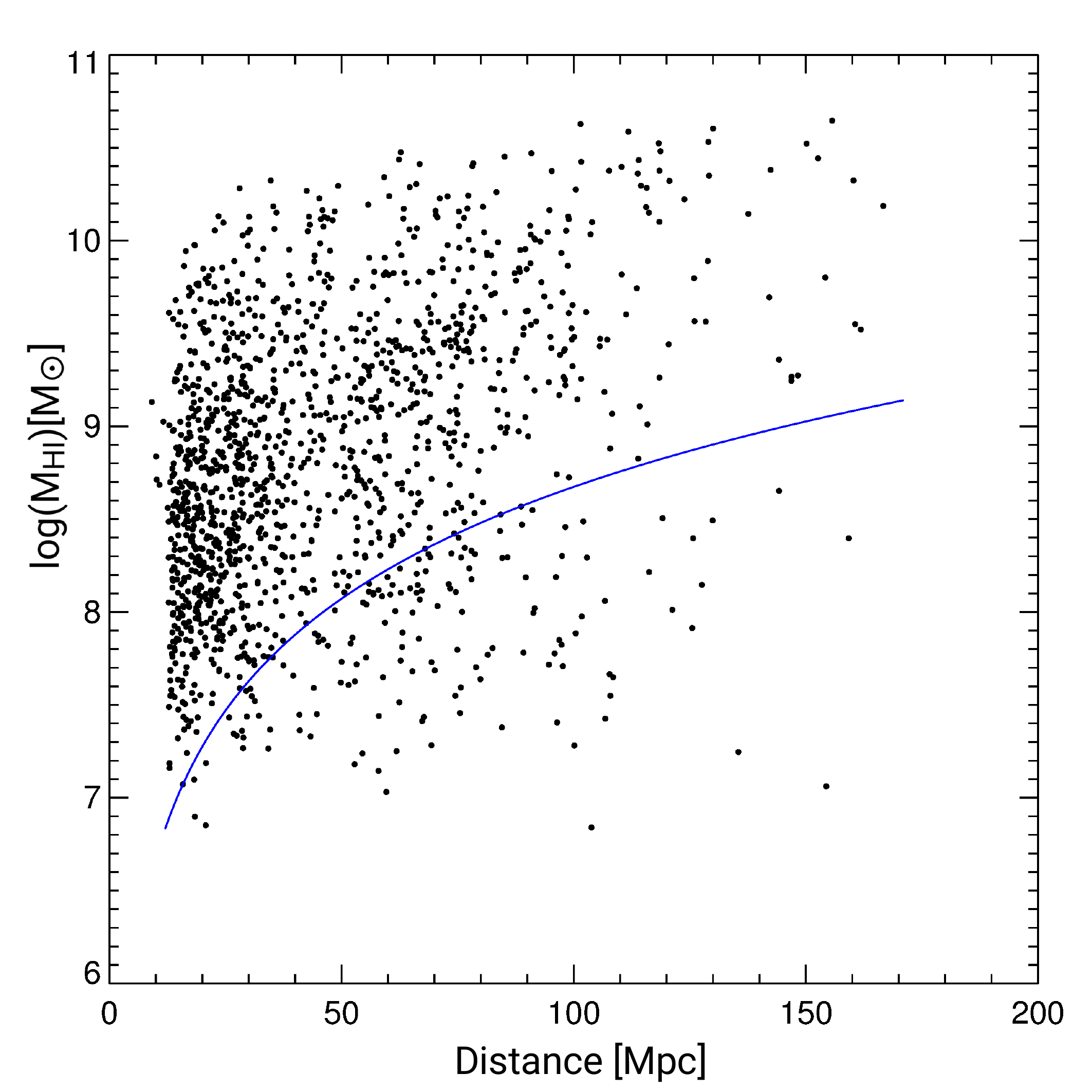}
\caption{\small Artificial log \HI\ masses in \Msun\ as a function of distance (in Mpc) for the NIBLES sources.  The blue line is the minimum detectable \HI\ mass based on an 0.2 \Jykms\ integrated line flux.}
\label{fig:MHI_dist}
\end{figure}

\begin{figure}
\centering
\includegraphics[width=8.5cm]{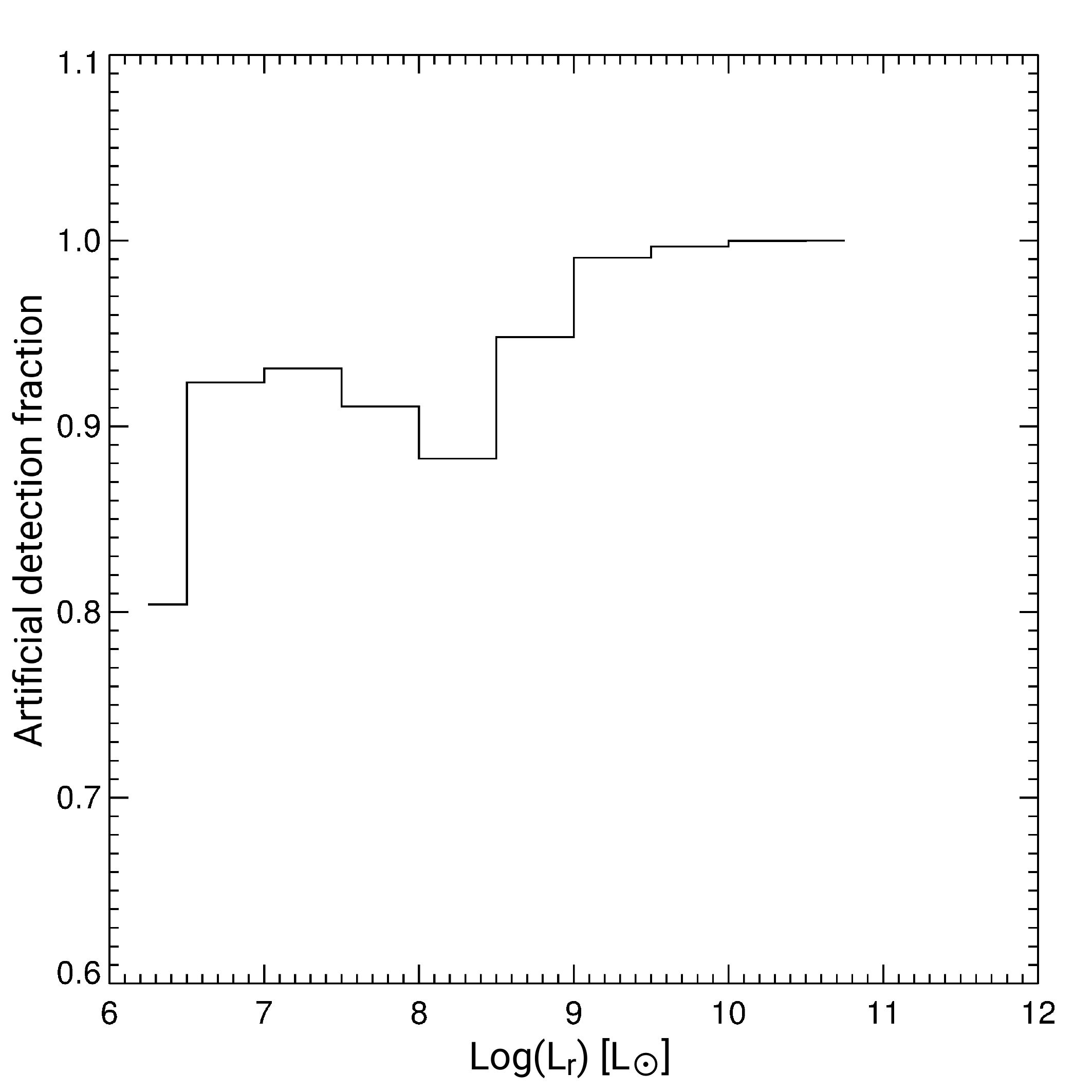}
\caption{\small Artificial detection fraction as a function of $r$-band luminosity for our artificial \HI\  masses.}
\label{fig:NRTdetfrac_fake}
\end{figure}

We then compare this minimum detectable \HI\ mass to the assigned artificial \HI\ masses based on the observed log(\MHI/\Lr) distribution shown in Fig. \ref{fig:MLratio_Lr}.  Fig. \ref{fig:MHI_dist} shows an example of our artificially generated \HI\ masses as a function of distance, with the blue line indicating our minimum detectable \HI\ mass.  The fraction of galaxies that lie above their corresponding minimum detectable mass in each luminosity bin gives us our artificial detection fractions.  We ran 100 iterations of the artificial mass distributions and used the mean values of each luminosity bin to derive our artificial detection fractions as a function of luminosity.  These detection fractions are shown in Fig. \ref{fig:NRTdetfrac_fake}.  Note that these fractions are only meant to illustrate the decreasing detection fractions of low mass galaxies as a function of luminosity due to distance effects alone, not to accurately describe the effects of \HI\ depletion in larger, more luminous galaxies.  There is roughly a 10\% decrease in detection fraction for the sources with log(\Lr) $\leq$ 8.25 compared to those with log(\Lr) $\geq$ 9.25.  This is similar to the approximately 10\% observed decrease in detection fraction discussed in Sect. \ref{sec:HI_sens_lim} as a function of decreasing luminosity.  
    
Since we can construct artificial \HI\ mass distributions for the NIBLES sample that show a similar decrease in detection fraction with decreasing luminosity as the observed detection fractions, we can conclude that it is likely that the approximately 10\% decrease in observed detection fractions is  due, at least in part, to distance effects only.  We can therefore correct for this in the BLF.
    
\end{appendix}
    
\end{document}